\documentclass[sigplan,nonacm]{acmart}

\usepackage{algorithm}

\newcommand{\myState}[1]{\State \scriptsize #1}
\newcommand{\myComment}[1]{\Comment{\tiny #1\scriptsize}}
\usepackage{xspace}

\usepackage{subfig}
\usepackage{algpseudocode}
\algnewcommand{\LineComment}[1]{\State \(\triangleright\) #1}
\newcommand{\myLineComment}[1]{\LineComment{\tiny #1\scriptsize}}

\usepackage[]{hyperref}

\newcommand{\ignore}[1]{}

\newcommand{\fig}[1]{Figure~\ref{#1}}
\newcommand{\sect}[1]{Section~\ref{#1}}

\newcommand{\algo}[1]{Algorithm~\ref{#1}}

\newcommand{\proposed}[0]{LazyDP\xspace}
\newcommand{\dpsgdf}[0]{DP-SGD(F)\xspace}
\newcommand{\lazydpwoans}[0]{LazyDP(w/o ANS)\xspace}
\newcommand{\lazydp}[0]{LazyDP\xspace}
\newcommand{\eana}[0]{EANA\xspace}
\newcommand{\sgd}[0]{SGD\xspace}

\newcommand\blfootnote[1]{%
\begingroup
\renewcommand\thefootnote{}\footnote{#1}%
\addtocounter{footnote}{-1}%
\endgroup
}

\makeatletter
\def\@titlefont{\LARGE \bfseries \par}
\makeatother

\begin{document}

\title{LazyDP: Co-Designing Algorithm-Software for Scalable Training of Differentially Private Recommendation Models}
\renewcommand{\shorttitle}{LazyDP: Co-Designing Algo-SW for Scalable Training of Differentially Private Rec. Models}

\author{Juntaek Lim}
\affiliation{%
    \institution{KAIST}
    \country{}
}\email{juntaek0425@kaist.ac.kr}

\author{Youngeun Kwon}
\affiliation{%
    \institution{KAIST}
    \country{}
}\email{yekwon@kaist.ac.kr}

\author{Ranggi Hwang}
\affiliation{%
    \institution{KAIST}
    \country{}
}\email{ranggi.hwang@kaist.ac.kr}

\author{Kiwan Maeng}
\affiliation{%
    \institution{Pennsylvania State University}
    \country{}
}
\email{kvm6242@psu.edu}

\author{G. Edward Suh}
\affiliation{%
    \institution{FAIR at Meta \\ Cornell University}
    \country{}
}
\email{edsuh@meta.com}

\author{Minsoo Rhu}
\affiliation{%
    \institution{KAIST}
    \country{}
}
\email{mrhu@kaist.ac.kr}
\renewcommand{\shortauthors}{J. Lim, Y. Kwon, R. Hwang, K. Maeng, G. E. Suh, and M. Rhu}
\begin{abstract}
Differential privacy (DP) is widely being employed in the
industry as a practical standard for privacy protection. While private training of computer vision or natural language processing applications has  been studied extensively, the computational challenges of  training of recommender systems (RecSys) with DP have not been explored. 
In this work, we first present our detailed characterization of private RecSys training using DP-SGD, root-causing its several performance bottlenecks. Specifically, we identify DP-SGD's noise sampling and noisy gradient update stage to suffer from a severe compute and memory bandwidth limitation, respectively, causing significant performance overhead in training private RecSys.
Based on these findings, we propose \proposed, an algorithm-software co-design that addresses the compute and memory challenges of training RecSys with DP-SGD. 
Compared to a state-of-the-art DP-SGD training system, we demonstrate that \proposed provides an average 119$\times$ training throughput improvement while also ensuring mathematically equivalent, differentially private RecSys models to be trained.
\end{abstract}

\maketitle
\pagestyle{plain}
\blfootnote{
This is the author preprint version of the work. The authoritative version will appear in the Proceedings of the $29^{\text{th}}$ ACM International Conference on Architectural Support for Programming Languages and Operating Systems (ASPLOS-29), 2024.
}

\section {Introduction}
\label{sect:intro}
With recent advances in deep neural networks (DNNs), hyperscalers leverage DNN-based recommender systems (RecSys) to power their  ads recommendation service~\cite{wideanddeep,netflix:rec,aibox, facebook_dlrm,dlrm:arch, microsoft_recsys,youtube_recsys}. The widespread adoption of RecSys, however, is raising serious concerns on protecting the \emph{privacy} of users. This is because the most popular machine learning (ML) models utilized in personalized ads are trained over private user data (e.g., websites a user has visited, items a user has purchased).
With increasing emphasis on privacy-preserving machine learning (PPML), \emph{differential privacy} (DP) is gaining momentum as a well accepted notion of privacy and is
widely being deployed in industrial applications, particularly for training ML algorithms requiring provable privacy guarantees~\cite{meta_fl_dp,apple,amazon,google}.

One of the most widely adopted \emph{private} ML training algorithm is DP-SGD (differentially private stochastic gradient descent~\cite{abadi}), which is an extension to  \emph{non-private} SGD~\cite{sgd}.  There are three important steps in DP-SGD that distinguish it from SGD: (1) unlike SGD which derives \emph{per-batch} gradient during backpropagation, DP-SGD requires the
derivation of \emph{per-example} gradient, (2) to limit the influence of each training example, the per-example gradient is norm-clipped when the L2 norm of the gradient exceeds a predefined threshold, and (3) to guarantee DP, the averaged gradient is added with \emph{Gaussian noise} before the gradient is used to update model weights (\sect{sect:sgd_vs_dpsgd}).
With the recent surge of interest in PPML, there has been several prior work exploring DP-SGD in terms of its privacy, model accuracy, and its training throughput~\cite{diva,reweighted,eana,fast_dp_sgd, amazon_mixed_dp,kurakin2022training, de2022unlocking,anil2021largescale, ponomareva-etal-2022-training-text, google, hoory2021learning, li2022large, yu2022differentially}. 
These studies, however, primarily focused on computer vision~\cite{amazon_mixed_dp,kurakin2022training, de2022unlocking} or natural language processing~\cite{anil2021largescale, ponomareva-etal-2022-training-text, google, hoory2021learning, li2022large, yu2022differentially}, so there is a dearth of prior work providing a systematic evaluation of DP-SGD's computational challenges for RecSys~\cite{eana,fast_dp_sgd}.

Given this landscape, an important motivation and  key contribution of
this work is the characterization and analysis on private RecSys training using DP-SGD, root-causing its bottlenecks from a systems perspective.
Modern RecSys employs categorical features to 
enhance accuracy, which is modeled using embedding layers. An embedding layer utilizes an \emph{embedding table} (which is an array of embedding vectors) and \emph{gathers} multiple
embedding vectors to translate categorical features into embeddings.
A unique property of embedding layers is that accesses to these  tables are extremely ``sparse'', e.g., embedding layer in MLPerf DLRM~\cite{mlperf} accesses only $0.03\%$ of the  table content per each training iteration. Because the embedding tables are also model weights subject for training, 
an embedding layer trained using non-private SGD involves a sparse \emph{gradient update} operation, targeting gradient updates only to the table locations storing the embeddings gathered during forward propagation.

Training embeddings with DP-SGD, on the other hand, requires the addition of Gaussian noise to \emph{all} the embeddings within the embedding table, regardless of whether they were subject for embedding gathers or not during forward propagation. This in turn transforms SGD's sparse gradient update into a dense \emph{``noisy'' gradient update} operation that modifies the \emph{entire} embedding table.
Our characterization uncovers the following two critical system-level challenges of DP-SGD's noisy gradient update operation (\sect{sect:characterization_noise}).

\begin{itemize}
\item {\bf Noise sampling.} Adding Gaussian noise to embedding vectors requires the sampling of a noise value from a normal distribution. 
Because noise sampling is required for every single embedding table entry and each sampling 
involves a series of trigonometric and logarithmic operations, noise sampling 
incurs \emph{high computational overheads}.

\item {\bf Noisy gradient update.} 
As the noisy gradient is a dense tensor sized identically to the embedding table, noisy gradient update 
exhibits a \emph{memory bandwidth limited behavior}. For large table sizes, which is common in modern RecSys models~\cite{yi2018factorized,mudigere2021high, aibox, isca2022:mudigere}, the noisy gradient update operation incurs severe performance overheads.

\end{itemize}

In this work, we present our algorithm-software co-design  \emph{\proposed} that enables scalable and high-throughput DP-SGD training with privacy guarantees for RecSys. \proposed can protect an arbitrary training example against an adversary who has access and control over the final trained model and all the other training data. While the adversary LazyDP assumes is slightly weaker than the original DP-SGD, our proposal provides the same DP-SGD level of privacy protection from most of the practical adversaries (\sect{sec:threat_model} details our assumed threat model). \proposed is based on the following key contributions that effectively addresses the compute (noise sampling) and memory (noisy gradient update) bottlenecks of private embedding layer training. 

\begin{itemize}
\item {\bf Lazy noise update.} 
Due to the sparse access nature of embedding gathers, the majority of embeddings are not accessed at any given training iteration, only updated using the noise but not the gradient. Generalizing this observation, if an embedding has not been accessed during the \emph{current} training iteration, it is most likely not going to be accessed again during the \emph{next} iteration (and the iteration after that),
 only going through a series of noise updates without any accesses.
Given such property, 
unless an embedding is scheduled to be accessed in its next training iteration, our proposal
 ``delays'' the noise update process for the  embeddings at a given training iteration.
  Such optimization helps alleviate the memory bandwidth bottleneck of noisy gradient updates as the majority of noise update operations can be skipped (delayed), significantly reducing memory traffic.
  To make sure our lazy noise update algorithm does not affect the final trained model, i.e., the privacy of baseline DP-SGD, \proposed ensures that all delayed noise updates are properly conducted \emph{before} an embedding is actually accessed (\sect{sect:proposed:algorithm}).

\item {\bf Aggregated noise sampling.} While lazily adding noise to the embedding table helps reduce memory traffic incurred with noise updates, the compute bottlenecks of the noise sampling itself still remain. This is because the embedding that delayed its noise addition over $N$ consecutive training iterations  must eventually be accumulated with
$N$ separate noise values, each sampled from a normal distribution and causing identical noise sampling overhead vs. baseline DP-SGD. LazyDP exploits the mathematical principles behind normally distributed random variables and proposes ``aggregated'' noise sampling, a novel noise sampling algorithm which dramatically reduces the compute overheads of noise sampling by orders of magnitude (\sect{sect:proposed:ans}).

\end{itemize}

Overall,  LazyDP  provides an average $119\times$ training speedup over representative RecSys models while guaranteeing mathematically equivalent, differentially private RecSys models to be trained vs. baseline DP-SGD.

\section{Background}
\label{sect:background}

\subsection{Recommendation Models}
\label{sect:recsys}

\begin{figure}[t!] \centering
	\includegraphics[width=0.47\textwidth]{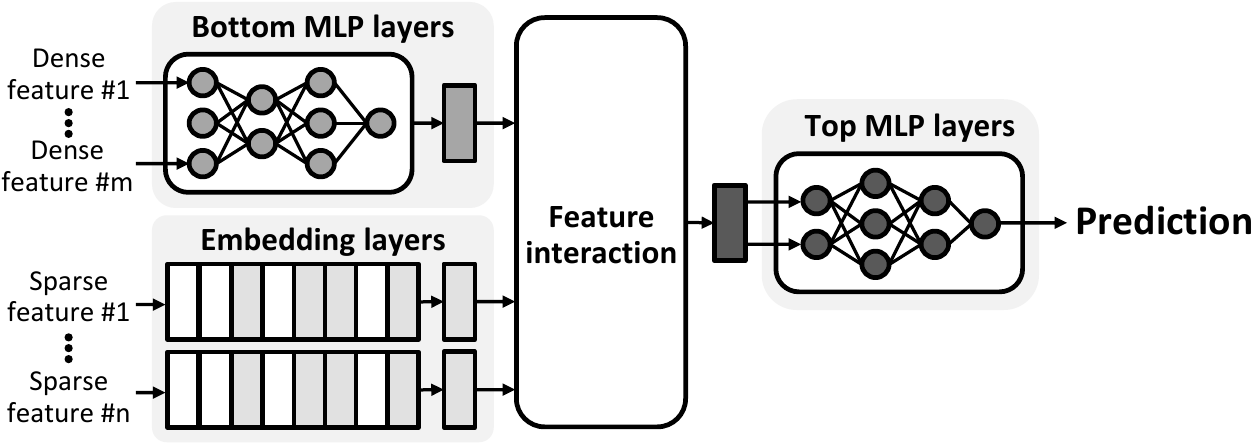}
\caption{
A RecSys model architecture using embedding layers.
}

\vspace{-0.5em}
\label{fig:recsys}
\end{figure}

Modern RecSys typically combines two key components, a sparse embedding layer and a dense DNN layer with multi-layer perceptions (MLP). Such model architecture enables  both sparse categorical features as well as dense continuous features to be captured, enhancing the model's accuracy. In order to handle sparse categorical features, the embedding layer utilizes \emph{embedding tables}
to convert a categorical feature into an \emph{embedding vector} (\fig{fig:recsys}).
An embedding table is an array of embedding vectors and is indexed using a sparse feature input, which is a discrete valued index ID. As there can be millions to billions of items that can fall under a feature category (e.g., all videos available on Netflix), an embedding table can be sized at several tens to hundreds of GBs, with recent large-scale RecSys even reaching TB-scale~\cite{yi2018factorized,mudigere2021high,aibox, isca2022:mudigere}. When predicting a RecSys model output, multiple embedding vectors can be \emph{gathered} from the embedding table, all of which are \emph{pooled} into a single vector using a reduction operation, e.g., element-wise addition or multiplication.

Embedding layers have two distinct characteristics compared to dense DNN layers. First, they exhibit extremely low compute intensity because both embedding gather and pooling operations are limited by \emph{memory bandwidth}. Secondly, embedding layers require \emph{high memory capacity} due to the sheer size of the embedding tables, which need to encompass all categorical information as embedding vectors.

\subsection{System Architecture for Training RecSys}
\label{sect:sysarch}

The size of embedding tables in recent RecSys reached several hundreds of GBs to even TB-scale.
As ML training is a throughput-oriented algorithm, 
system architectures for RecSys training typically employ throughput-optimized GPUs which employ bandwidth-optimized memory based on stacked DRAM (e.g., HBM). These high-bandwidth memory solutions, however, come with limited capacity, only providing several tens of GBs which makes it impossible to store the memory hungry embedding tables locally within the GPU.

To this end, state-of-the-art system architectures for training RecSys commonly employ a CPU-GPU based system~\cite{aibox,fb:zion,scratchpipe,tensorcasting, isca2022:mudigere}. Under such system design, the CPU is provisioned with a large pool of capacity-optimized DIMMs (e.g., LR-DIMM) to locally store large embedding tables, enabling the training of embedding layers using the CPU. The training of compute-intensive DNN layers is handed over to the GPU to best exploit its high compute and memory throughput. In the rest of this paper, we assume such CPU-GPU based system as the baseline system to train RecSys.

\subsection{Differential Privacy}
\label{sect:dp}

As ML applications have become more popular and widespread, the need to protect a user's data privacy has also grown significantly. This is because the training dataset used in ML is typically collected using crowdsourcing and may include sensitive private information of users. Given such vulnerability, several prior work  demonstrated that an adversary that has only black-box access to the ML model can restore individual training examples through training data extraction, leaking private information~\cite{carlini_llm, carlini_diffusion, haim_reconstruction, plug_n_play, fredrikson_drug, fredrikson_face, plug_in_inversion, carlini_chatgpt}.

Given this landscape, differential privacy (DP) has received a lot of attention in recent years as a well-accepted notion of privacy thanks to its
strong and mathematically rigorous guarantees on privacy protection~\cite{dwork2006dp}. For an ML model to be differentially private,
 the trained model should be indistinguishable 
whether or not a particular training example was utilized to train the ML model, allowing the privacy
of individual training examples to be protected with DP. 
In \cite{abadi}, Abadi et al. presented their seminal work on training DNN based ML algorithms using DP, aka DP-SGD. In the following subsection, we explain both non-private SGD and private DP-SGD training and discuss their key differences. For the interested readers, we refer to 
\cite{abadi, dwork2006dp, dwork2008differential, dwork2014algorithmic} for a
more detailed discussion on the mathematical foundation behind DP.

\begin{figure}[t!]
    \centering
    \vspace{-0.5em}
    \subfloat[]{\includegraphics[width=0.26\textwidth]{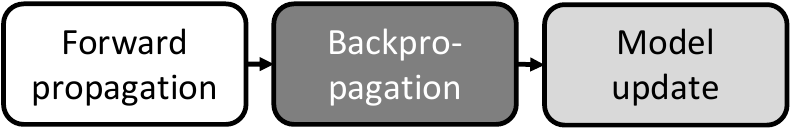}\label{fig:NN_training}}
    \hfill
    \hspace{1.0em}
    \centering
    \subfloat[]{\includegraphics[width=0.19\textwidth]{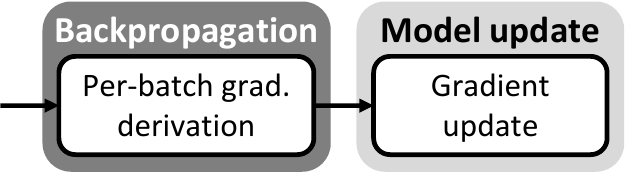}\label{fig:NN_training_SGD}}
    \hfill
    \centering
    \vspace{-0.5    em}
    \subfloat[]{\includegraphics[width=0.47\textwidth]{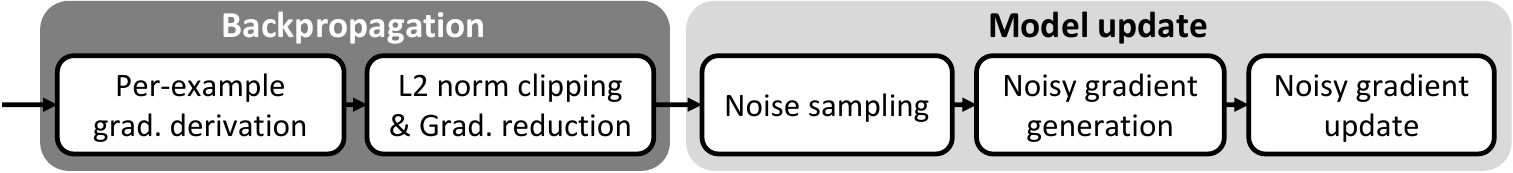}\label{fig:NN_training_DPSGD}}
    \caption{
    Non-private SGD vs. private SGD. (a) SGD and DP-SGD both go through the same set of operations during all of forward propagation and the activation gradient derivation of backpropagation. Key differences between the backpropagation and model update of (b) SGD and (c) DP-SGD include the L2 norm clipping, noise sampling, and updating the model weights with the noisy gradient.
    }
    \label{fig:sgd_vs_dpsgd}
\vspace{-0.5em}
\end{figure}

\subsection{Non-private SGD vs. Private DP-SGD Training}
\label{sect:sgd_vs_dpsgd}

Training an ML model requires updating its model weights  through forward, backward propagation (i.e., backpropagation), and model update. Both  SGD and DP-SGD involve an identical set of layer execution during forward propagation (\fig{fig:sgd_vs_dpsgd}(a)). 
As for backpropagation, SGD and DP-SGD go through the same set of procedures to derive each layer's \emph{activation gradient} but significantly differ in the way the \emph{weight gradient} is derived. We summarize their differences below.

{\bf Non-private SGD.}
To update the ML model weights, the weight gradient must first be derived at a size identical to the target model size, which is subsequently utilized by the optimizer to train the model.
A non-private SGD does this by \emph{aggregating} the weight gradients derived for individual input examples that constitute an input mini-batch, effectively \emph{reducing} them down into a single set of weight gradient (\fig{fig:sgd_vs_dpsgd}(b)), the size of which becomes identical to the model weight size. In this paper, we refer to SGD's weight gradient as the \emph{per-batch} weight gradient to differentiate it from DP-SGD's \emph{per-example} gradient as explained below.

{\bf Private DP-SGD.} A key ingredient of DP-SGD~\cite{abadi} is to \emph{clip} and add \emph{noise} to the weight gradients during updates. As illustrated in \fig{fig:sgd_vs_dpsgd}(c), 
rather than computing an averaged per-batch gradient, DP-SGD first derives the gradient for each \emph{individual example} within a mini-batch.
These \emph{per-example} gradients are first clipped using a predetermined threshold, using L2 norm. After clipping, all the per-example gradients are averaged, and Gaussian noise is added. The final \emph{``noisy'' gradient} is used to update the model.

\subsection{Related Work}
\label{sect:related}

{\bf Fast and memory-efficient DP-SGD.}
A key challenge with DP-SGD is that the derivation of per-example weight gradients requires separate memory allocations for each per-example weight gradient across all the layers, causing a memory capacity bottleneck (e.g., a mini-batch size of $N$ requires $N$
times larger memory allocations for storing weight gradient tensors in DP-SGD).  To alleviate DP-SGD's memory
capacity problem, Lee et al.~\cite{reweighted} proposed \emph{reweighted} DP-SGD (henceforth referred to as DP-SGD(R)) which reduces the memory
allocation size of weight gradients at the expense of  additional computation overheads. Specifically, 
 DP-SGD(R) goes through the weight gradient derivation computation \emph{twice}, (1) first computing the per-example weight gradient but
 only to derive the L2 norm clipping threshold values, which enables DP-SGD(R) to not have to incur high memory allocation overheads of the original DP-SGD, and then (2) utilizing these L2 norm values, DP-SGD(R) conducts a \emph{reweighted} version of per-batch weight gradient derivation to derive the final aggregated weight gradient values, i.e., per-batch weight gradient.
 
 While DP-SGD(R)'s need to go over both per-example and per-batch  weight gradient derivation may seem to lead to worse performance than the original DP-SGD, DP-SGD(R) opens up unique opportunities to \emph{fuse} several key bottleneck layers, achieving higher performance while also incurring smaller memory allocation size than DP-SGD~\cite{fast_dp_sgd,diva}. Additionally, the output model trained with DP-SGD(R) is mathematically identical to the original memory hungry DP-SGD algorithm.
 Because of DP-SGD(R)'s higher performance and smaller memory allocation while also guaranteeing the original DP-SGD's privacy protection, we consider DP-SGD(R) as a stronger baseline DP-SGD algorithm for characterization in \sect{sect:characterization}. Recent work by Denison et al.~\cite{fast_dp_sgd} proposed a performance-optimized version of DP-SGD(R) (referred to as DP-SGD(F) in the rest of this paper) which is based on observations that RecSys only consists of embedding layers and linear MLP layers, allowing for the efficient estimation of the L2 norm of per-example gradient using standard backpropagation, without the need for deriving per-example weight gradient.
 As we detail in \sect{sect:characterization_end_to_end}, DP-SGD(F) consistently provides higher performance than DP-SGD(R), so we assume DP-SGD(F) as the baseline DP-SGD algorithm to compare against \proposed in our final evaluation in \sect{sect:eval}.

 Other than these closely related prior work, DiVa~\cite{diva} is a recently proposed accelerator architecture for high-performance DP-SGD training. Unlike \proposed which improves the performance of  private sparse embedding layer training, DiVa
 focuses on optimizing the performance of \emph{non}-embedding layers, only targeting computer vision and natural language processing ML algorithms. As such, 
 DiVa suffers from the exact same compute and memory bottlenecks of RecSys training that GPU systems currently encounter upon, one which we detail in 
 \sect{sect:characterization}.
In general, the key contribution of \proposed is orthogonal to DiVa as our proposal complements
DiVa's limitation by broadening its applicability for RecSys.

{\bf DP-SGD for training RecSys.} Denison et al.~\cite{fast_dp_sgd}  presents a privacy-utility trade-off analysis of employing DP-SGD for training real world ads dataset, demonstrating that DP-SGD can provide both privacy and good model accuracy for RecSys. Similar to \proposed, EANA~\cite{eana} presents an alternative DP-SGD algorithm to alleviate the compute and memory overheads of training RecSys embedding tables, improving overall training throughput.
Unlike \proposed, however, EANA's privacy protection is fundamentally weaker because it modifies the original DP-SGD algorithm to add noise \emph{only} to the accessed embedding vectors, weakening DP-SGD's ability to anonymize individual training example.
As an intuitive example, EANA never adds noise to an embedding vector if it has never been accessed, which will directly leak the fact that no user data contains the corresponding feature -- the original DP-SGD algorithm does not leak such information because it adds noise to \emph{all} the embedding vectors regardless of whether they have been accessed. \proposed does not alter the mathematical foundation behind the baseline DP-SGD algorithm and enjoys the same level of privacy guarantee, while significantly alleviating its performance overhead. In \sect{sect:eval:eana}, we compare \proposed against EANA to demonstrate \proposed's merits.
\section{Threat Model}
\label{sec:threat_model}

LazyDP protects an arbitrary training example against an adversary who has access and control over (1) the final trained model and (2) all the other training data except for the specific example under attack. The adversary that LazyDP assumes, while slightly weaker than what the original DP-SGD can handle, is still highly relevant to practical scenarios.

The original DP-SGD can protect against an adversary with an access and control over (1) the final trained model, (2)  all the other training data, and additionally, (3) \emph{all the intermediate states} of the model (i.e., all the gradients) during the training process~\cite{abadi}. This original threat model of DP-SGD is often considered excessively strong~\cite{adversary_instantiation}, as many real-world adversaries can only access the final trained model, either directly or through an inference API, but cannot access the intermediate gradients.
For example, recent real-world attacks that successfully extracted training data from a released GPT-2 model~\cite{carlini_llm}, released Stable Diffusion model~\cite{carlini_diffusion}, and by querying ChatGPT API~\cite{carlini_chatgpt}, were all done only with the access to the final trained model and not the intermediate gradients.
When considering these practical adversaries, LazyDP provides the exact same privacy guarantee as the original DP-SGD.
The same weaker but practical threat model was assumed in many other prior work as well~\cite{adversary_instantiation, eana, dp_iteration, milad_attack, label_only_attack, chuan_attack} including EANA~\cite{eana}.
In fact, the reason for the overly strong threat model of DP-SGD (that does not match real-world attackers' capability) is due to the current limitation in privacy accounting. So far, there is no known mathematical technique to directly analyze privacy against the more practical adversaries that cannot access the intermediate gradients~\cite{adversary_instantiation}.

Admittedly, there are cases where the adversary can access the gradients. Such cases include federated learning~\cite{fl}, where the gradients of the individual trainer are sent to an untrusted server. LazyDP can also be vulnerable in special cases where the parameter server in a distributed training is considered untrusted~\cite{laoram} or uses a specialized hardware~\cite{switchml, parameterhub} that may be hacked.
LazyDP cannot provide protection that matches the original DP-SGD in these cases.

\begin{figure}[t!] \centering
\includegraphics[width=0.47\textwidth]{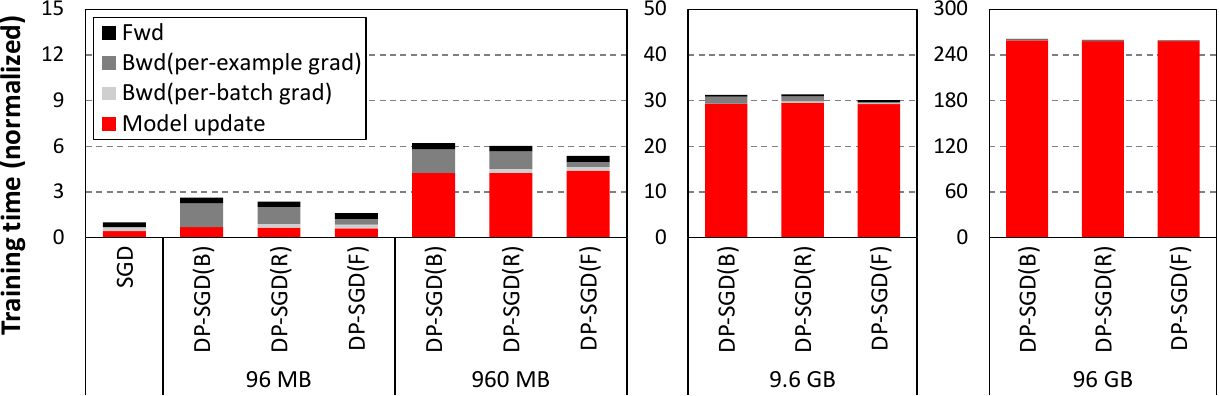}
\caption{
Breakdown of SGD and DP-SGD's training time into key stages of forward and backward propagation. SGD's training time remains almost constant regardless of the table size, so this figure only shows a single SGD data point under the default 96 GB for brevity. All data points are normalized to this SGD (the leftmost bar).
}
\label{fig:motivation_latency}
\vspace{-0.7em}
\end{figure}

\section{Workload Characterization on Private RecSys Training with DP-SGD}
\label{sect:characterization}

This section takes a top-down approach in characterizing the 
computational challenges of training RecSys with DP-SGD. 
We use DLRM (deep learning recommendation model~\cite{facebook_dlrm}) from MLPerf~\cite{mlperf} as our 
target RecSys workload whose 
default model size is $96$ GB. 
In order to demonstrate the effect embedding table size has on DP-SGD's training time, we scale down the size of this default model by reducing the number of embedding table entries, from 10$\times$$\downarrow$ ($9.6$ GB) to 1000$\times$$\downarrow$ ($96$ MB).  RecSys training is modeled using PyTorch Opacus~\cite{opacus} which we utilize to evaluate not only the original DP-SGD algorithm (denoted DP-SGD(B)) but also its  performance-optimized versions as discussed in \sect{sect:related}, namely DP-SGD(R)~\cite{reweighted} and DP-SGD(F)~\cite{fast_dp_sgd}. \sect{sect:methodology} further details our evaluation methodology.

\subsection{Breakdown of End-to-End Training Time} 
\label{sect:characterization_end_to_end}

We start our characterization by first trying to gain a high-level understanding on what impact DP training has for RecSys.
\fig{fig:motivation_latency} compares the end-to-end training time of non-private SGD vs. three private DP-SGD designs (DP-SGD(B,R,F)), uncovering several key observations as detailed below.

\begin{figure}[t!]
    \centering
    \vspace{-0.5em}
    \subfloat[]{\includegraphics[height=0.15\textwidth]{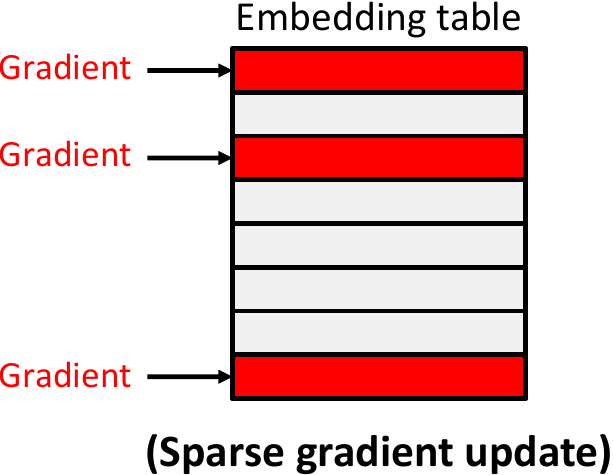}\label{fig:table_update_sgd}}
    \hspace{1.2em}
    \centering
    \subfloat[]{\includegraphics[height=0.15\textwidth]{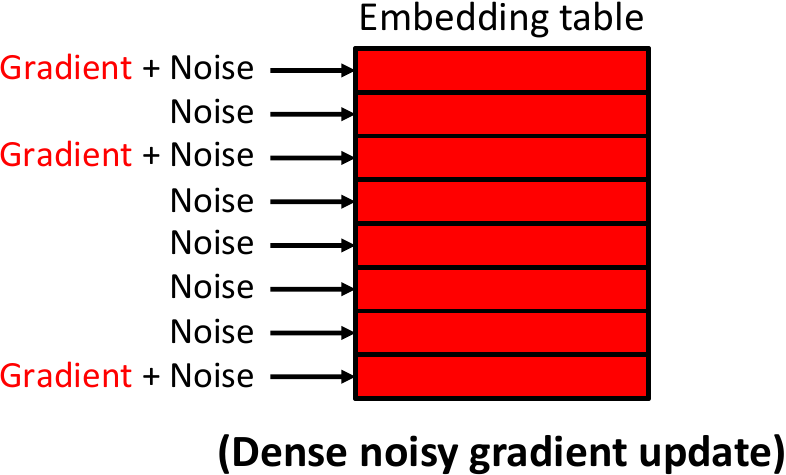}\label{fig:table_update_dpsgd}}
    \caption{
    Training an embedding table using (a) SGD and (b) DP-SGD. Example assumes a pooling value of $3$. (a) SGD only incurs $3$ table reads and $3$ table writes during
    forward propagation and model update,
    respectively, exhibiting sparse table accesses. (b) In contrast, DP-SGD requires an additional $8$ noise write operations on top of SGD's default $3$ table read and $3$ table write operations, collectively invoking a dense noisy gradient update during model update.
    }
    \vspace{-0.7em}
    \label{fig:table_update_sgd_vs_dpsgd}
\end{figure}

In general, the training time of SGD remains almost constant regardless of the table size, so we only show a single SGD data point in \fig{fig:motivation_latency} for brevity. In contrast, all three DP-SGD designs experience an almost linear increase in latency as table size increases.
The reason why DP-SGD experiences aggravated training performance is as follows.  
A non-private SGD training exhibits a \emph{sparse} embedding table access pattern, only touching a small \emph{fixed} number of embeddings (determined by the pooling value and batch size) during both forward propagation (embedding gather) and model update (gradient update), regardless of table size. In contrast, training an embedding table with DP-SGD requires the addition of Gaussian noise to \emph{all} the embeddings within the embedding table during model update, regardless of whether they were targeted for an embedding gather operation during forward propagation. In effect, an SGD's sparse gradient update turns into a dense \emph{noisy gradient} update with DP-SGD that updates the \emph{entire} embedding table entries, severely pressurizing both compute and memory of the training system (\fig{fig:table_update_sgd_vs_dpsgd}). 

Consequently, the larger the embedding table size becomes, the more embedding vectors DP-SGD must access to go over the computations required to add noise to the embeddings. 
An important point to note in \fig{fig:motivation_latency} is that the performance difference among the three DP-SGD designs gradually diminishes as the model size increases, e.g., the most performance-efficient DP-SGD(F) provides $1.5\times$ higher performance than DP-SGD(R) when the table size is the smallest at $96$ MB, but the performance gap becomes less than $0.3\%$ under the default size at $96$ GB.
This is because the performance overhead of updating the embedding table with both noise and gradient (included as part of the \emph{model update} stage in \fig{fig:motivation_latency}) grows proportional to the table size, washing out any performance improvement DP-SGD(R) and DP-SGD(F) provides on top of the baseline DP-SGD(B). In other words, prior work on performance-efficient DP-SGD (DP-SGD(F)), while effective in reducing DP-SGD(B)'s backpropagation latency to derive gradients, is not able to fundamentally address the bottlenecks incurred in training the embedding tables with DP.

\emph{ Key takeaways: Training embedding tables with DP-SGD requires every single table entry to be added with Gaussian noise, invoking a dense noisy gradient update to the entire table and causing a severe performance overhead. 
}

\begin{figure}[t!] \centering
\includegraphics[width=0.47\textwidth]{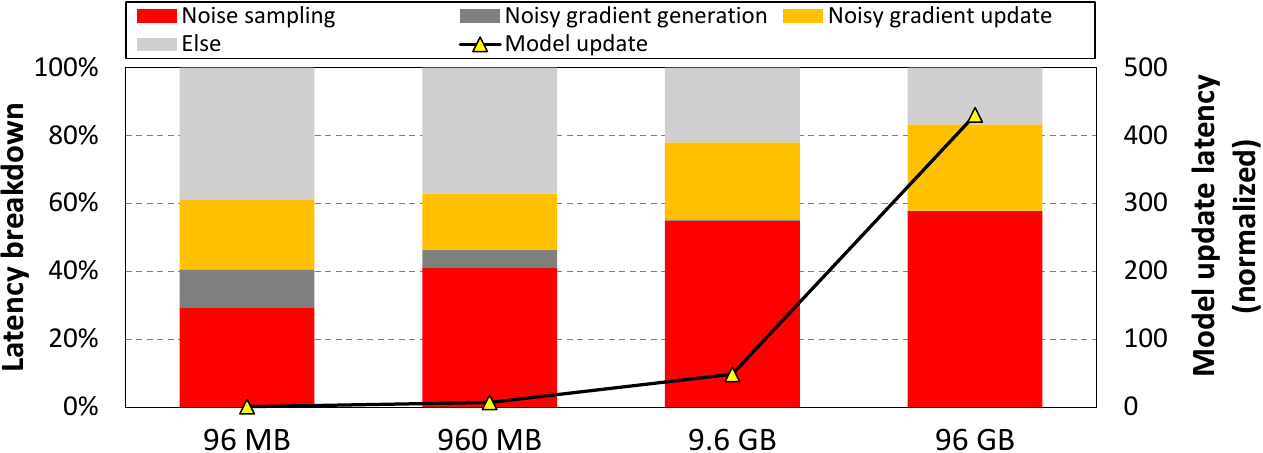}
\caption{
Latency breakdown (left axis) of the model update stage in \fig{fig:motivation_latency}. The right axis shows the latency in model update (normalized to $96$ MB), the value of which grows as model size increases.
}
\label{fig:latency_breakdown_update}
\vspace{-0.5em}
\end{figure}

\subsection{Analysis on the Model  Update Stage in Private Embedding Table Training}
\label{sect:characterization_update}

The previous subsection identified the model  update stage as becoming a critical performance bottleneck in training private RecSys. In  \fig{fig:latency_breakdown_update}, we further break down the latency of the model update stage (red in \fig{fig:motivation_latency}) into four parts: noise sampling, noisy gradient generation, noisy gradient update, and others. 
We emphasize that, rather than utilizing PyTorch's built-in functions \emph{as-is} for our characterization, \emph{we heavily optimize and tune the performance of this bottleneck stage to construct a strong, competitive baseline DP-SGD}. More concretely, \emph{our optimized version of the model update stage is $8.2\times$ faster than the baseline implementation using the original built-in PyTorch functions}, reaching \emph{$81\%$ of the maximum possible AVX performance achievable for this stage}. We further analyze our optimized DP-SGD's model  update stage in the following subsection (\sect{sect:characterization_noise}).

As depicted in \fig{fig:latency_breakdown_update}, the noise sampling and noisy gradient update account for a significant fraction of the bottlenecked model update stage, with its proportion rapidly increasing as the embedding table size is increased. For instance, the aggregate latency of noise sampling and noisy gradient update accounts for $83.1\%$ of the model update stage under the default $96$ GB model size, consuming $82.8\%$ of its end-to-end training time and becoming the two most significant performance limiters. As 
previously pointed out in \sect{sect:characterization_end_to_end}, the amount of memory traffic to \emph{update} the embedding table using the noisy gradient (and accordingly, the overhead of \emph{generating} each noise value to update each embedding vector) grows proportional to the embedding table size, so we can infer that these two bottlenecks will only get worse for future RecSys models with even larger table sizes~\cite{isca2022:mudigere,aibox}.

\emph{ Key takeaways: Even with a highly optimized software implementation of DP-SGD, we root-cause its two most significant performance bottlenecks as the step that generates noise and the noisy gradient update step, the overhead of which becomes even worse for future large-scale RecSys. 
}

\subsection{Root-causing the Key Challenges behind DP-SGD's Noise Sampling and Noisy Gradient Update}
\label{sect:characterization_noise}

Since our optimized version of model update stage achieved $8.2\times$ speedup vs. baseline PyTorch's implementation, a natural question arises: Can the two most bottlenecked stages, i.e., noise sampling and noisy gradient update, be optimized even further and remove their performance bottlenecks completely? Conversely, are noise sampling and noisy gradient update stages fundamental bottlenecks that require alternative measures to address their performance overheads?

To answer these questions, we conducted an in-depth examination of the key primitives that constitute both noise sampling and noisy gradient update.
In PyTorch, Gaussian noise sampling function is implemented in the torch.normal() API, which generates a tensor of a specified size containing independent random numbers following a normal distribution. 
To achieve high computational efficiency, this API is implemented based on the Box-Muller algorithm~\cite{box_muller_wiki}, which is well-known to provide high performance random number sampling to processors with vector units.
Careful examination of this API (which provides our noise sampling functionality) revealed that its implementation is dominated by conducting a series of trigonometric and logarithmic computations using AVX (Advanced Vector Extensions~\cite{avx2}) instructions. More concretely, we observe that the majority of torch.normal()'s execution time is dominated
by executing a series of (1) AVX load instruction to retrieve an input vector, (2) $101$ AVX compute instructions for trigonometric/logarithmic/other operations, and finally (3) AVX store instruction to write back the result, exhibiting a highly \emph{compute-bound} behavior. The noisy gradient update stage, on the other hand, exhibits a \emph{memory bandwidth limited} behavior as it is primarily a data streaming operation, i.e., each data element is loaded from memory, multiplied by a learning rate, and then added with the existing model weight, requiring only two computations for each loaded data element while streaming a large amount of data.

\begin{figure}[t!] \centering
    \includegraphics[width=0.47\textwidth]{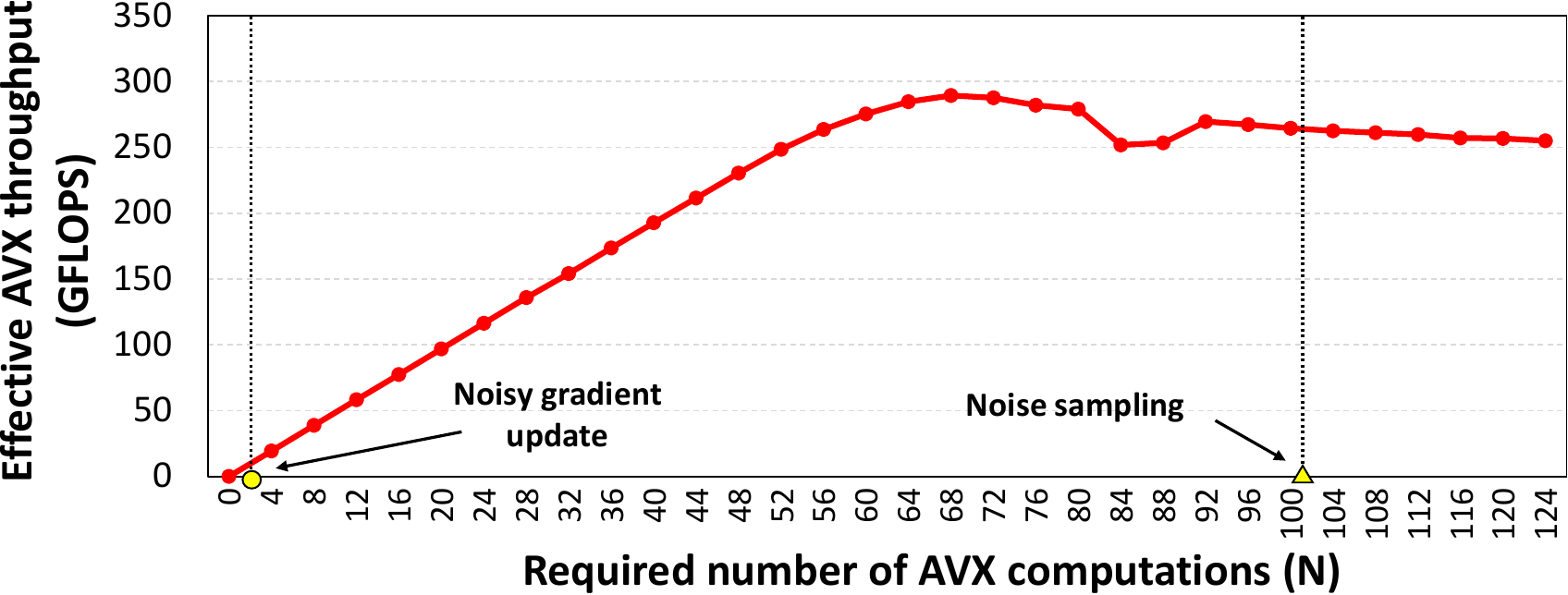}\label{fig:motivation_avx}
\caption{
Effective AVX throughput as a function of the number of AVX computations ($N$) conducted over a single loaded vector. As depicted, our implementation of noise sampling (corresponding to the data point at $N$=$101$) exhibits a compute-bound behavior and reaches $81\%$ of the maximum AVX performance.
}
\label{fig:microbenchmark}
\vspace{-0.5em}
\end{figure}

To better illustrate the compute-bound (noise sampling) and memory-bound (noisy gradient update) nature of model update stage, we design a microbenchmark that conducts (1) an AVX vector load from memory, (2) performs $N$ consecutive AVX computations over the loaded vector, and then (3) writes back the resulting vector into memory using AVX store. We then sweep the value of $N$ and measure the effective AVX throughput, the result of which is illustrated in \fig{fig:microbenchmark}. As shown, when the number of $N$ (number of AVX computations performed over the loaded vector) is small, the benchmark exhibits a memory-bound behavior whereas for large $N$ values, it becomes compute-bound. The compute-bound noise sampling stage, with its large $N$ value of $101$, achieves an effective AVX throughput of $215$ GFLOPS and reaches $81\%$ of the maximum possible AVX performance. As for the memory-bound noisy gradient update, which corresponds to a small $N$ value of $2$, it is able to reap out $85.5\%$ of theoretical memory bandwidth and $99.8\%$ of the maximum possible AVX performance. Overall, these results confirm that the noise sampling and noisy gradient update operators we utilize as our baseline DP-SGD(B,R,F) leave little performance left on the table  because they already reach close to the optimal performance under the training system's available compute and memory throughput constraints. 

\emph{ Key takeaways: DP-SGD's performance limiting noise sampling and noisy gradient update operations already reach close to its optimal performance, suggesting that alternative measures must be devised to address the performance bottlenecks of private RecSys training.
}

\section{LazyDP: An Algorithm-Software Co-Design for Differentially Private RecSys Training}
\label{sect:proposed}

In this section, we present our \proposed system, an algorithm-software co-design for private RecSys training. \proposed enables high throughput DP-SGD training while also ensuring that the trained RecSys model brings the same level of privacy protection provided with baseline DP-SGD against the practical adversary discussed in Section~\ref{sec:threat_model}. \sect{sect:proposed:observation} first introduces the design principles behind \proposed and the key observations that underpin our proposal. We then detail  \proposed's lazy noise update algorithm co-designed with our aggregated noise sampling technique 
for scalable DP-SGD training (\sect{sect:proposed:implementation}). Lastly, \sect{sect:proposed:software_arch} describes  \proposed
's software architecture and its user interface.

\vspace{-0.5em}
\subsection{Design Principles and Key Observations}
\label{sect:proposed:observation}
\begin{figure}[t!]%
\vspace{-0.5em}
\centering
\subfloat[SGD]{%
\label{fig:delayed_update_a}%
\includegraphics[height=1.96in]{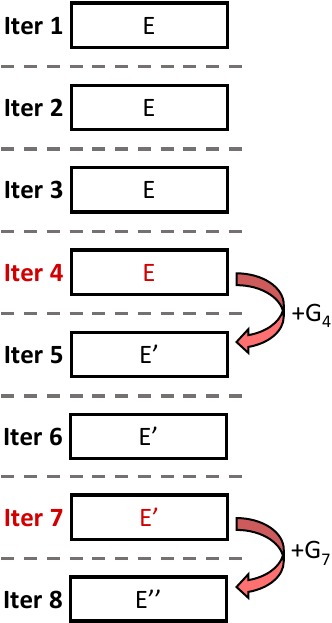}}%
\hspace{0.4em}
\subfloat[DP-SGD]{%
\label{fig:delayed_update_b}%
\includegraphics[height=1.96in]{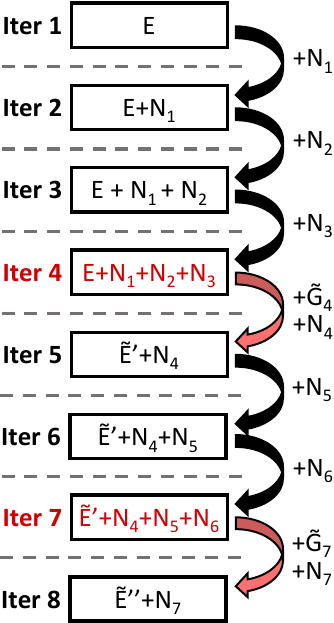}}%
\hspace{0.4em}
\subfloat[\proposed]{%
\label{fig:delayed_update_c}%
\includegraphics[height=1.96in]{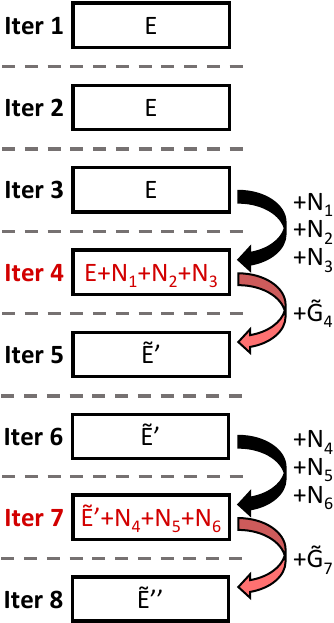}}%
\caption{An example embedding's gradient and noise update process using (a) SGD, (b) conventional DP-SGD, and (c) \proposed's lazy noise update algorithm. The red arrow indicates an embedding update with gradient $G_i$, a value derived during the $i$-th training iteration. The black arrow indicates an update with the noise value of N$_{i}$, which is the noise that should be updated in iteration $i$.
The tilde symbol represents a  value derived with DP in mind (e.g., $\tilde{G_i}$, $\tilde{E}'$ and $\tilde{E}''$ are values derived with noise values considered in its derivation). 
}
\label{fig:delayed_update}%
\end{figure}

An important observation behind \proposed is that delaying the noise update process (which we refer to as \emph{lazy noise update}) of DP-SGD can significantly improve computational efficiency while still ensuring that the privacy of final training outcome (i.e., the trained RecSys model) is not affected. 
We use \fig{fig:delayed_update} as a running example to motivate our lazy noise update algorithm, describing key differences between SGD, DP-SGD,
and our proposed LazyDP.

In this example, we show the set of gradient and/or noise updates a \emph{single} embedding vector experiences during the course of eight consecutive training iterations. Due to embedding table's sparse access pattern, the example assumes that an embedding gather operation happens to this particular embedding vector only during the $4^{th}$ and the $7^{th}$ training iteration. Recall that SGD only updates the weights of those embedding vectors that are gathered during forward propagation. As such, the non-private SGD updates this embedding vector only during the iterations that it actually accessed, i.e., G$_{4}$ and G$_{7}$ is derived during the $4^{th}$ and the $7^{th}$ iteration to update this embedding vector (\fig{fig:delayed_update}(a)). The baseline DP-SGD training, on the other hand, adds noise to each embedding vector in every training iteration, significantly amplifying memory read and write traffic compared to SGD (all N$_{i}$ that update noise to this embedding, \fig{fig:delayed_update}(b)).

\proposed's design is based on the key observation that, as long as we make sure that any delayed noise updates are conducted \emph{before} the actual embedding access occurs, the exact timing of \emph{when} those delayed noise updates were performed have no impact on the gradient values derived ($\tilde{G_4}$ and $\tilde{G_7}$). Consider the embedding vector update that happens on the $4^{th}$ iteration in \fig{fig:delayed_update}(c). Here, \proposed delays the noise updates until the $3^{rd}$ training iteration (which immediately precedes the iteration that actually accesses the target embedding, i.e., the $4^{th}$ iteration) and 
conducts the noise update in aggregation by adding (N$_{1}$+N$_{2}$+N$_{3}$) altogether.
This allows gradient $\tilde{G_4}$ to be derived in the $4^{th}$ iteration using the embedding with a value of (E+N$_{1}$+N$_{2}$+N$_{3}$), the same value that was  used to derive $\tilde{G_4}$ under baseline DP-SGD. Unlike DP-SGD, however, the number of memory reads to fetch the embedding vectors (as well as memory writes to store the updated embeddings) is significantly reduced as 
the three consecutive noise updates of (N$_{1}$+N$_{2}$+N$_{3}$) are performed all at once, rather than going over each noise update separately in each training iteration (\fig{fig:delayed_update}(b)). Generalizing this behavior across the entire embedding table, we can infer that the number of (delayed) noise updates invoked with \proposed is primarily determined by the embedding table pooling value (i.e., number of embedding gathers per table) per each iteration.
This is because \proposed's lazy noise update only occurs for those embeddings that will be gathered during the next training iteration, the number of which is solely determined by the table's pooling value but not the table size. Such property helps \proposed significantly reduce memory traffic than DP-SGD, which must always update the entire table in every training iteration. Consequently, \proposed's reduction in memory read/write traffic helps address the bottleneck in the memory-bound noisy gradient update operation (\fig{fig:latency_breakdown_update}), improving training throughput. Nonetheless, it is important to point out that the overhead of \emph{noise sampling} itself (e.g., generating N$_{1}$, N$_{2}$, and N$_{3}$) has not changed between \fig{fig:delayed_update}(b) and \fig{fig:delayed_update}(c). We discuss how the bottlenecks of noise sampling are addressed in \sect{sect:proposed:ans} using our aggregated noise sampling algorithm.

Note that, in order to perform the lazy noise updates right before an embedding vector is about to be accessed, \proposed must be able to \emph{predict} which embedding vectors will be accessed in the subsequent training iteration. Another key observation we make is that the nature of RecSys training dataset enables us to \emph{precisely} determine which embedding vectors will be gathered at what time, which \proposed can utilize to determine when to initiate lazy noise update. In RecSys, the training dataset describes which embedding table entries to gather from during forward propagation, not just for the current iteration but also for all future iterations as well. \proposed is designed to prefetch the next training iteration's training mini-batch in order to analyze which embedding vectors will be accessed in the subsequent iteration and determine whether to initiate lazy noise update or not. Since \proposed requires visibility on which embeddings will be accessed in one subsequent training iteration,  prefetching a single mini-batch in advance is sufficient to implement our proposal. In the following subsection, we detail the implementation of our lazy noise update and aggregated noise sampling algorithm.

\subsection{Implementation}
\label{sect:proposed:implementation}

\subsubsection{Lazy Noise Update Algorithm} 
\label{sect:proposed:algorithm}

In \algo{algo:lazydp}, we provide a pseudo-code of \proposed's lazy
noise update algorithm.
For brevity, the pseudo-code omits the model update process for the dense MLP layers because both \dpsgdf and \proposed apply the identical DP protection for MLP layers. Below we walk through this pseudo-code to describe its working mechanism and pinpoint to where \proposed's performance improvement comes from.

\textbf{Line 1-2} introduces the $HistoryTable$, a data structure that tracks the number of delayed noise updates for each embedding vector. A naive implementation of the $HistoryTable$ will be to simply count the number of delayed noise updates each embedding vector is pending, incrementing the per-embedding counter value whenever that embedding is not accessed in each training iteration. 
Considering the significant access sparsity of embedding tables, however, such naive implementation will lead to significant memory write traffic to update the $HistoryTable$. \proposed instead employs the $HistoryTable$ to track the most recent training iteration ID number that performed a lazy noise update to each embedding vector. By calculating the distance between the current training iteration and the ID value stored within the $HistoryTable$ (line 14), we can derive the number of delayed noise updates. As the value of $HistoryTable$ is only updated for the \emph{sparsely} accessed embedding table entries (line 15), the performance overhead of maintaining and updating the $HistoryTable$ is very low.

\begin{algorithm}[t!]
\caption{Lazy Noise Update Algorithm.}
\label{algo:lazydp}
\begin{algorithmic}[1]
\scriptsize

\renewcommand{\algorithmicrequire}{\textbf{(Parameters)}}
\Require model weight: $\theta$, max clipping factor: $C$, noise multiplier: $\sigma$, number of iterations: $N$, training batch size: $B$, learning rate: $\eta$, embedding dimension: $dim$, number of embedding vectors in the embedding table: $E$
\vspace{0.5em}
\myLineComment{Data structure recording the latest noise updated iter. for each embedding vector}
\myState  $HistoryTable \leftarrow \Call{Zeros}{E}$
\myComment{Initialize to an array of length $E$ containing zeros}

\vspace{0.4em}
\myLineComment{Data structure to store the current and next input mini-batches}
\myState  $InputQueue \leftarrow \Call{Queue}{size = 2}$ \myComment{Contains two consecutive input mini-bathes}
\myState $InputQueue.push(\Call{GetNextMinibatch}{})$  \myComment{Load the first training mini-batch}

\vspace{0.4em}
\For {$iter \leftarrow$ $1$ \textbf{to} $N$}
    \myState $InputQueue.push(\Call{GetNextMinibatch}{})$  \myComment{Load a new training mini-batch}

    \vspace{0.4em}
    \myLineComment{Forward and backward propagation is done identically to standard DP-SGD}
    \myState $loss \leftarrow \Call{ForwardPropagation}{InputQueue.head(), \theta}$
    \myState $gradient \leftarrow \Call{Backpropagation}{loss, \theta, C}$ \myComment{Training with a grad clipping}
    \vspace{0.4em}

    \myLineComment{Retrieve the next iteration's input mini-batch from the tail of the InputQueue}
    \myState $next\_accesses \leftarrow InputQueue.tail().sparse\_inputs$
    \For {\textbf{each} $idx$ \textbf{in} $next\_accesses$} \myComment{Derive number of delayed noise updates}
        \myState $delays[idx] \leftarrow iter - HistoryTable[idx]$
        \myState $HistoryTable[idx] \leftarrow iter$ \myComment{Renew the value of the \textit{HistoryTable}}
    \EndFor
        
    \vspace{0.4em}    
    \myLineComment{Noise sampling using next mini-batch  and number of delayed noise updates}
    \myState $noise \leftarrow \Call{NoiseSampling}{next\_accesses, delays, B, C, dim}$
    
    \vspace{0.4em}
    \myLineComment{Generate noisy gradient}
    \myState $noisy\_gradient \leftarrow gradient + noise$ \myComment{Merge sparse gradient and sparse noise}
    
    \vspace{0.4em}
    \myLineComment{Model update}
    \For {\textbf{each} $idx$ \textbf{in} $noisy\_gradient.indices$}
        \myLineComment{Weight update}
        \myState $\theta_{sparse}[idx] \leftarrow \theta_{sparse}[idx] - \eta \times noisy\_gradient[idx]$
    \EndFor
    \vspace{0.4em}
    \myState $InputQueue.pop()$ \myComment{Pop the head of the InputQueue}
\EndFor
\vspace{0.8em}
\Procedure{NoiseSampling}{$next\_accesses, delays, B, C, dim$}
    \For {\textbf{each} $idx$ \textbf{in} $next\_accesses$}
        \If {$ANS$ $disabled$}
            \myLineComment{Accumulate Gaussian noises of size $dim$, variance of $\sigma^2 C^2$ iteratively}
            \myState $noise[idx] \leftarrow \Call{Zeros}{dim}$    \myComment{$noise[i]$: noise vector for $i^{th}$ embedding}
            \For {$n \leftarrow$ $1$ \textbf{to} $delays[idx]$}
                \myState $noise[idx] \leftarrow noise[idx] + \frac{1}{B}\times \Call{GaussianNoise}{\sigma^2 C^2, dim}$ 
            \EndFor
        \Else   \myComment{ANS enabled}
            \myLineComment{Generate an aggregated noise all at once using the ANS algorithm}
            \myState $noise[idx] \leftarrow \frac{1}{B}\times \Call{GaussianNoise}{delays[idx]\times\sigma^2 C^2, dim}$ 
        \EndIf
    \EndFor
    \myState \Return $noise$
\EndProcedure

\end{algorithmic}
\end{algorithm}

\textbf{Line 3-5} introduces the $InputQueue$, which enables \proposed to preview the next input mini-batch. For the purpose of identifying which embeddings will require lazy noise updates, a queue with a size of two is sufficient, storing two consecutive input mini-batches. When the training system is bootstrapped, the first training mini-batch is retrieved from the data storage system (line 5). During the main training iteration loop (line 6-27), \proposed fetches the next mini-batch and reuses the previous iteration’s next mini-batch as the current mini-batch, having visibility into \emph{both} current and next input mini-batches while only fetching a \emph{single} mini-batch each iteration (line 7) -- identical to baseline SGD and DP-SGD.

\textbf{Line 6-27} explains the key operations undertaken during the main training loop. As described in line 8-10, \proposed requires no changes to forward and backpropagation, only changing the way the gradient and noise are used to update the embedding table.
Line 11-27 describes the important modifications required for lazy noise update, which performs noise sampling, noisy gradient generation, and noisy gradient update for model training.
Because lazy noise update targets embeddings that will be accessed in the next iteration, the tail of the $InputQueue$ is used (line 12) to identify the embedding vectors that will lazily be updated with noise.
We then use the $HistoryTable$ to 
calculate 
the number of delayed noise updates to perform for the target embedding vectors (line 13-16). The next process is the noise sampling stage (line 18) which samples the delayed amount of noise to add and then accumulate to generate the final noise vector. Depending on whether \proposed's aggregated noise sampling is employed or not (line 28-40), noise sampling can incur significantly different computational overheads, the details of which are discussed in the \sect{sect:proposed:ans}.

After noise sampling, \proposed generates the noisy gradient by merging the sampled noise with the gradient (line 19-20). The noisy gradient is then utilized to update the model based on the optimizer function (line 22-25). The performance gain of the lazy noise update comes from this phase. 
In baseline DP-SGD, the size of the noisy gradient is equivalent to the size of the entire embedding table so updating the model exhibits a memory bandwidth limited behavior. With our \proposed, however, 
the size of the noisy gradient is identical to the size of the embedding vectors accessed in the current and next iterations (line 20). Given the sparse access pattern of embedding tables, \proposed's noisy gradient becomes orders of magnitude smaller in size than DP-SGD's dense noisy gradient, providing significant reduction in the memory traffic for model update.

\begin{figure}[t!] \centering
\includegraphics[width=0.47\textwidth]{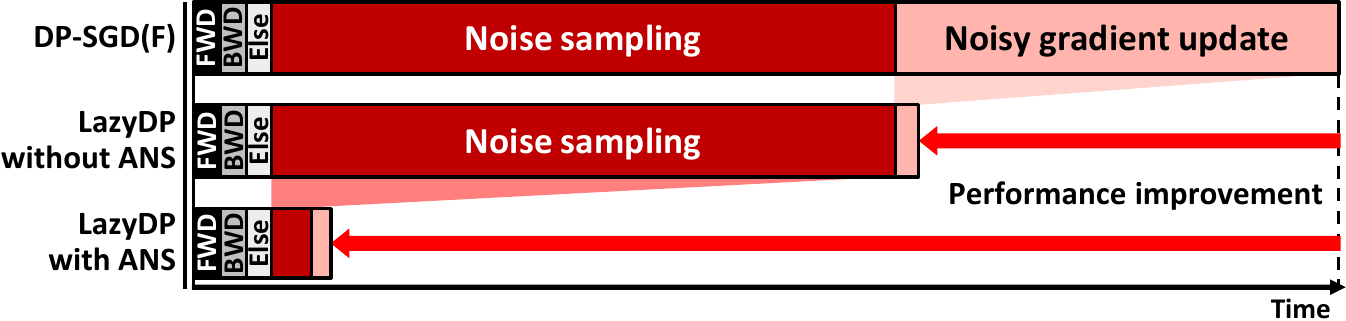}
\caption{\proposed's effect on latency reduction.}
\vspace{-0.5em}
\label{fig:proposed_visualization}
\end{figure}

\subsubsection{Aggregated Noise Sampling Algorithm}
\label{sect:proposed:ans}

Although lazy noise update alleviates the memory bottlenecks of noisy gradient update, the computational challenge of noise sampling still remains (\fig{fig:proposed_visualization}). This is because the total number of Gaussian noise to generate and accumulate into accessed embeddings has not changed vs. baseline DP-SGD, rendering noise sampling to still cause a performance bottleneck. We present our \emph{aggregated noise sampling} (ANS) algorithm, which leverages the mathematical property of normally distributed random variables to substantially reduce the number of noise sampling. In probability theory, it is well known
the summation of two or more random variables sampled from a normal distribution follows another normal distribution, but one with a different mean and variance~\cite{sum_of_random}. 
The following theorem describes this mathematical property.
\label{theorem:ans}
\begin{theorem} [\bf{Sum of i.i.d. normal random variables~\cite{sum_of_random}}]
Let $X_1, X_2, \dots, X_n$ be independent and identically distributed (i.i.d.) Gaussian random variables with a mean value of 0 and a variance of $\sigma^2$. Then, the summation of these random variables, denoted by $Y$, follows a Gaussian distribution with a mean of 0 and a variance of $n\sigma^2$.
Specifically, we have:
\[
X_i \sim N(0, \sigma^2) \quad \text{for } i = 1, 2, 
 \dots, n,
\]
where all $X_i$ are independent variables following Gaussian distribution with a mean of 0 and a variance of $\sigma^2$.
The summation of these independent variables is given by:
\[
Y = X_1 + X_2 + \dots + X_n, \hspace{0.5em} Y \sim N(0, n\sigma^2).
\]
\end{theorem}

In effect, what this theorem implies is that we can replace the summation of multiple independently sampled Gaussian noise into
a \emph{single} Gaussian noise sampled from a normal distribution with a larger variance. Coming back to our lazy noise update algorithm in \algo{algo:lazydp}, without ANS (line 30), the compute-bound noise sampling operation must be invoked for each delayed noise update (e.g., $3$ separate noise sampling operations to derive N$_{1}$, N$_{2}$, and N$_{3}$ in \fig{fig:delayed_update}(c)), causing significant computation overheads.
However, with the ANS algorithm, we can substitute multiple noise sampling operations (line 33-35) with a just a single $GaussianNoise$ sampling (line 38) from a normal distribution with a variance of ($delay\times\sigma^2 C^2$), instead of $\sigma^2 C^2$, significantly reducing the overhead of noise sampling (e.g., a single noise sampling operation derives N$_{1}$+N$_{2}$+N$_{3}$ in \fig{fig:delayed_update}(c)).

\subsection{Software Architecture}
\label{sect:proposed:software_arch}

\begin{figure}[t!]
    \centering
        \subfloat[]{\includegraphics[width=0.47\textwidth]{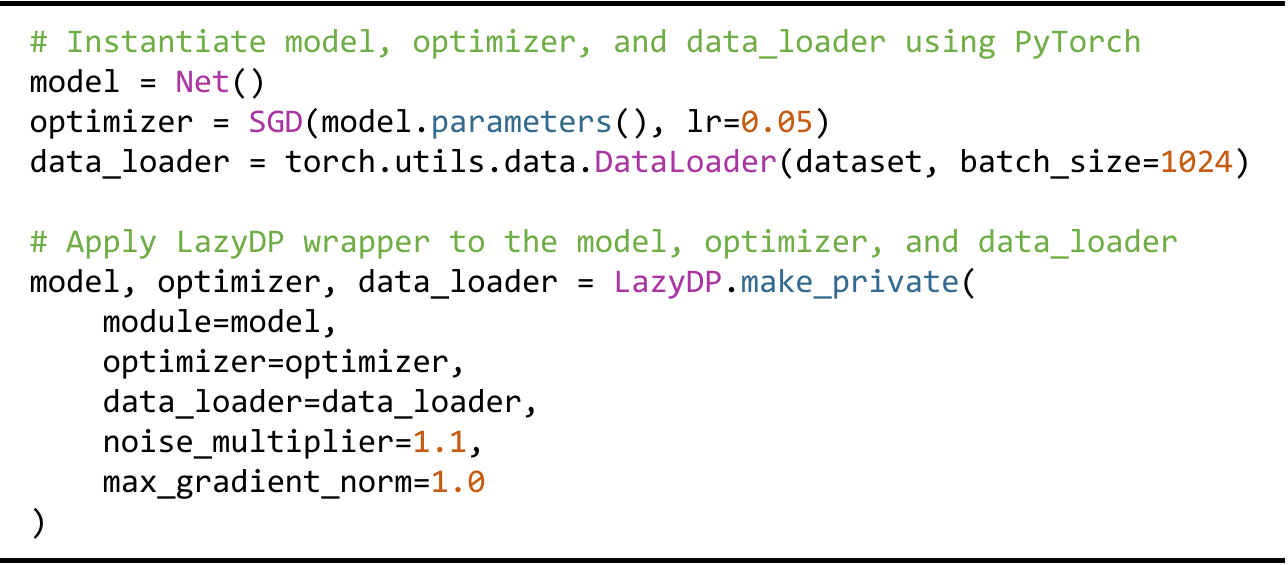}\label{fig:software_arch_code}}
    \hfill
    \subfloat[]{\includegraphics[width=0.47\textwidth]{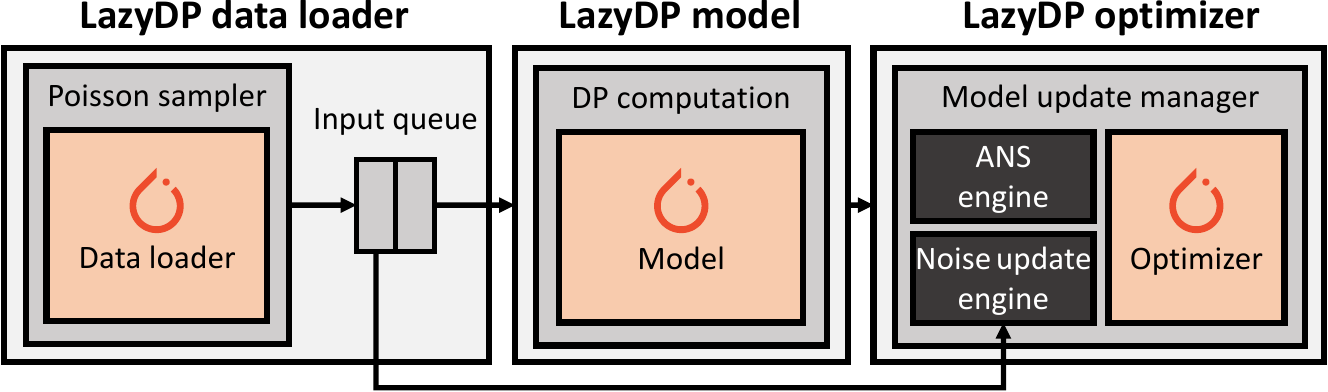}\label{fig:software_arch_diagram}}
    \caption{LazyDP's (a) user interface and (b) software architecture.}
    \label{fig:software_arch}
\vspace{-0.5em}
\end{figure}

\textbf{User interface.} \proposed is designed as a software plug-in to PyTorch Opacus~\cite{opacus}, extending the capabilities of existing ML frameworks to seamlessly support scalable DP-SGD training for RecSys. \fig{fig:software_arch}(a) shows the user interface provided with LazyDP, which transforms existing PyTorch model, optimizer, and data$\_$loader instances into LazyDP-enabled instances.
The wrapper API is also designed to accept various training hyperparameters to explore models with different model accuracy, privacy budget, and training performance. 

\textbf{Software system.}
The key components of \proposed are implemented as part of the backend software runtime system of PyTorch Opacus. \fig{fig:software_arch}(b) provides a high-level overview of the LazyDP software architecture, which wraps around PyTorch's data$\_$loader, model, and optimizer instances to implement our proposal.
We primarily utilize the DP-enabled data loader and model instances from PyTorch Opacus as-is to provide the same key features required for DP-SGD training, e.g., Poisson sampling based training dataset retrieval (data loader instance) and per-example gradient derivation (model instance). To provide visibility into both the current and the next training iteration's mini-batch inputs, however, \proposed's data loader instance is augmented with a two-entry input queue, which is forwarded to \proposed's model and optimizer instances for DP-based gradient derivation and model update, respectively. 
The \proposed optimizer instance incorporates our key optimizations, the \emph{ANS engine} and the \emph{noise update engine}, which implements the aggregated noise sampling and the lazy noise update algorithm, respectively.

{\bf Putting everything together}, the synergistic combination of the noise update engine and the ANS engine elegantly addresses the two most severe performance bottlenecks of private RecSys training, the compute-bound noise sampling and the memory-bound noisy gradient update. It is important to note that \emph{ANS builds on top of \proposed's ability to postpone noise sampling and noisy gradient update operations, a feature enabled by our lazy noise update algorithm}, highlighting the importance of \proposed's algorithm-software co-design.
\section{Methodology}
\label{sect:methodology}

{\bf System configuration.}
All our evaluations are conducted on a server containing NVIDIA V100 GPU (32 GB HBM2, 900 GB/sec bandwidth) and Intel Xeon E5-2698v4 CPU (256 GB of DDR4 DRAM, 68 GB/sec bandwidth) connected with PCIe 3.0 (16 GB/sec bandwidth). 
We assume a hybrid CPU-GPU system (\sect{sect:sysarch}) focusing on large-scale RecSys with several hundreds of GBs of embedding tables~\cite{aibox,tensorcasting, scratchpipe, ttrec}, where the CPU is used to train the sparse embedding layers and the GPU for the dense DNN layers.

{\bf Software.} We use PyTorch (v.1.12.1)~\cite{torch} and its Opacus~\cite{opacus} extensions for modeling the key functionalities required for DP-SGD, e.g., per-example gradient derivation, gradient clipping, Gaussian noise sampling, etc. However, as we detailed in \sect{sect:characterization_update}, the built-in functions that come with PyTorch Opacus as-is exhibited sub-optimal performance. For a conservative evaluation of \proposed, we heavily optimize the performance of the key bottleneck layers in baseline DP-SGD by using Intel TBB~\cite{tbb} and OpenMP~\cite{openmp} to optimally tune the way it leverages data-level parallelism (vectorized execution) and thread-level parallelism (multi-threading), achieving $13.4\times$ higher performance than the built-in PyTorch implementations and reaching $81\%$ of the ideal performance achievable (see \sect{sect:characterization_noise}). This performance optimized version of DP-SGD is used in all of our evaluations reported in this paper (the characterization study in \sect{sect:characterization} and the evaluation section in \sect{sect:eval}). Note that, because DP-SGD(F)~\cite{fast_dp_sgd} consistently exhibited higher performance than DP-SGD(R)~\cite{reweighted}, we only show the results using DP-SGD(F) when comparing against \proposed in \sect{sect:eval} for brevity.

{\bf Benchmarks.} 
We use the open-sourced DLRM~\cite{facebook_dlrm} and follow the MLPerf (v2.1)~\cite{mlperf} recommendation training benchmark's configuration to model our default RecSys model architecture, which uses $8$ MLP layers and $26$ embedding layers with a $128$-dimensional embedding, resulting in a total model size of 96 GB. The embedding table access patterns are drawn from a uniform distribution. In \sect{sect:eval:sensitivity}, 
we study the sensitivity of \proposed
when deviating from this default configuration.
\section {Evaluation} 
\label{sect:eval}

\begin{figure}[t!] \centering
\includegraphics[width=0.47\textwidth]{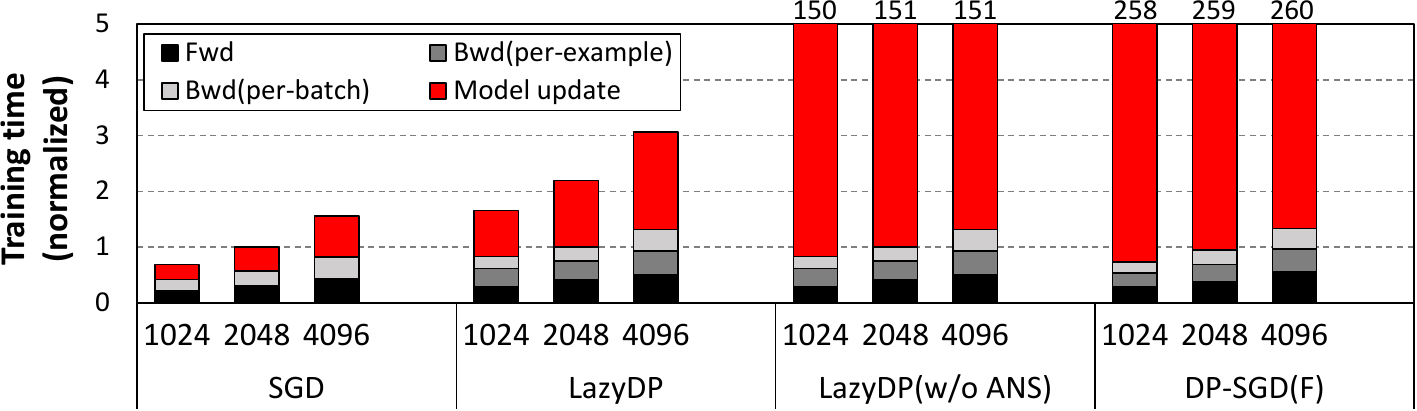}
\caption{End-to-end training time, normalized to SGD trained with mini-batch size 2,048.}
\label{fig:eval_performance}
\vspace{-0.5em}
\end{figure}

\begin{figure}[t!] \centering
\includegraphics[width=0.47\textwidth]{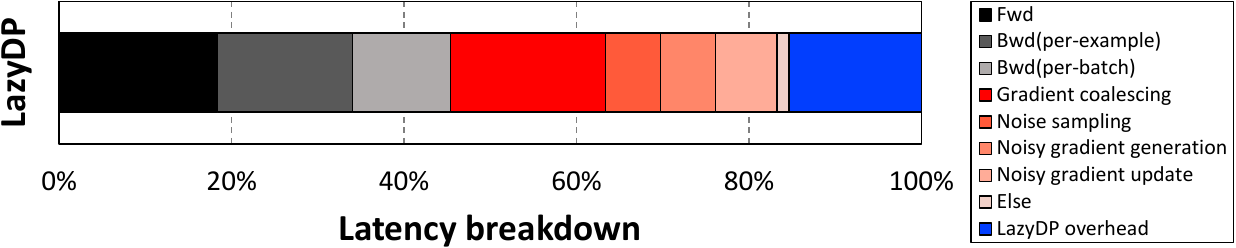}
\caption{Breakdown of LazyDP's training time when trained with mini-batch size 2,048.}
\label{fig:lazydp_latency}
\end{figure}

In this section, we first evaluate the performance and energy-efficiency of \proposed (\sect{sect:eval:performance}). We then discuss \proposed's implementation overhead (\sect{sect:eval:implementation_overhead}) and its sensitivity to different RecSys configurations (\sect{sect:eval:sensitivity}). Finally, \proposed is compared against \eana~\cite{eana}, a prior work that provides high-performance private RecSys training at the cost of reduced privacy guarantee (\sect{sect:eval:eana}).

\subsection{Performance and Energy-Efficiency}
\label{sect:eval:performance}

{\bf Performance.} In \fig{fig:eval_performance}, we compare the training time of non-private SGD and the private DP-SGD(F) and \proposed over three different training mini-batch sizes (1,024, 2,048, and 4,096). To demonstrate the effectiveness of our ANS optimization, we show both \proposed \emph{without} ANS (denoted \lazydpwoans) and \proposed with all of its optimizations incorporated (denoted \proposed). As discussed in \sect{sect:characterization}, training with \dpsgdf incurs significant latency overheads during the noise sampling and noise gradient update process, causing a 
$166-375\times$ performance slowdown vs. non-private \sgd. \lazydpwoans helps alleviate the bottleneck incurred in noisy gradient update, so it provides an average $72\%$ speedup on top of \dpsgdf. However, the performance overhead of noise sampling still remains with \lazydpwoans which renders its performance to still fall behind \sgd by $97-218\times$. \proposed with ANS holistically addresses all the critical bottlenecks of \dpsgdf and provides $85-155\times$ speedup, bringing down private DP-SGD’s training time to be on par with the non-private SGD. Specifically, our \proposed only incurs $1.96\times$ to $2.42\times$ slowdown over the non-private SGD, enabling private RecSys training to become practical for ML practitioners.

\begin{figure}[t!] \centering
\includegraphics[width=0.47\textwidth]{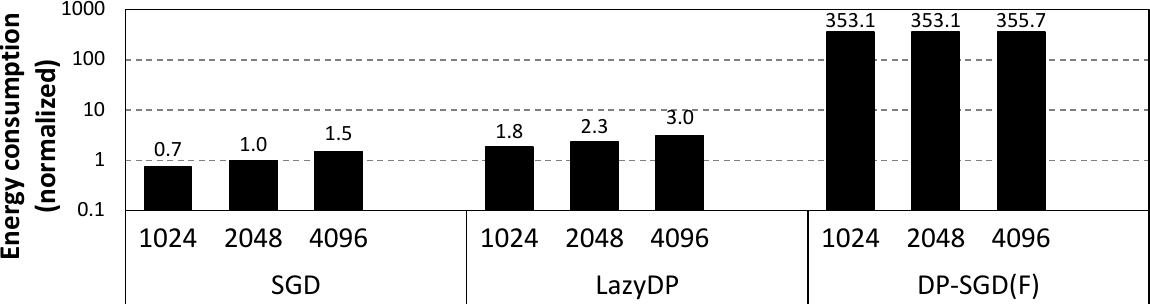}
\caption{Energy consumption (normalized to SGD, batch: 2,048).}
\label{fig:eval_energy}
\vspace{-1em}
\end{figure}

{\bf Latency breakdown.} To better illustrate where \proposed's speedup comes from, we show in \fig{fig:lazydp_latency} a latency breakdown of \lazydp under the default configuration trained with mini-batch size 2,048. Here, we further divide \proposed's model update stage (red bar in \fig{fig:eval_performance}) into finer granularity where the conventional operations conducted with DP-SGD training (e.g., noise sampling, noisy gradient generation, $\ldots$) are colored in reds and the pure \proposed-introduced latency overhead is colored in blue. Because \proposed's lazy noise update algorithm (\algo{algo:lazydp}) and ANS (\sect{sect:proposed:ans}) effectively reduce the latency of noise sampling and noisy gradient update operations by $1,081\times$ and $418\times$, respectively, no single operation is left as a major bottleneck anymore. The latency overheads of \proposed (the blue bar) include operations such as updating the \emph{HistoryTable}, identifying which embeddings to initiate lazy noise updates, and others, but all of these only account for $15\%$ of the training time of \proposed. Specifically, this $15\%$ latency overhead of \proposed is further broken down into three distinct components: (1) removing duplicated embedding indices among the embeddings accessed next, (2) reading the \emph{HistoryTable} and calculating the standard deviation for ANS, and (3) updating the 
 \emph{HistoryTable}, each component accounting for 61\%, 22\%, and 17\% to \proposed's overhead, respectively.

{\bf Energy-efficiency.} \proposed is a software-based solution that works seamlessly on top of existing systems. Consequently, the reduction in training time  directly translates into improved energy-efficiency. 
We utilize Intel's \texttt{pcm-power}~\cite{intel_pcm} and NVIDIA’s \texttt{nvidia-smi}~\cite{nvidia_smi} to measure
the power consumption of CPU and GPU, respectively, multiplying the aggregated CPU-GPU power number with its corresponding training time to calculate each design’s energy consumption. As shown in \fig{fig:eval_energy}, \proposed provides significant improvements in energy-efficiency by reducing energy consumption by an average $155\times$ against \dpsgdf.

\subsection{Implementation Overhead}
\label{sect:eval:implementation_overhead}

As discussed in \sect{sect:proposed:implementation}, our LazyDP requires additional metadata such as (1) the delayed input data queue that provides visibility into the next training iteration’s embedding table accesses, and (2) the $HistoryTable$ that tracks the number of delayed noise updates. As the input data queue only requires storing one additional training mini-batch, it incurs 213 KB (mini-batch size $\times$ number of embedding tables $\times$ average lookups per embedding table $\times$ 4 Bytes) of additional storage overhead. As for the $HistoryTable$, for our default 96 GB model configuration, it consumes an additional $751$ MB (total number of embedding vectors $\times$ 4 Bytes) of memory, which is less than $1\%$ of the total model size.

\begin{figure}[t!]
    \centering
    \subfloat[]{\includegraphics[width=0.47\textwidth]{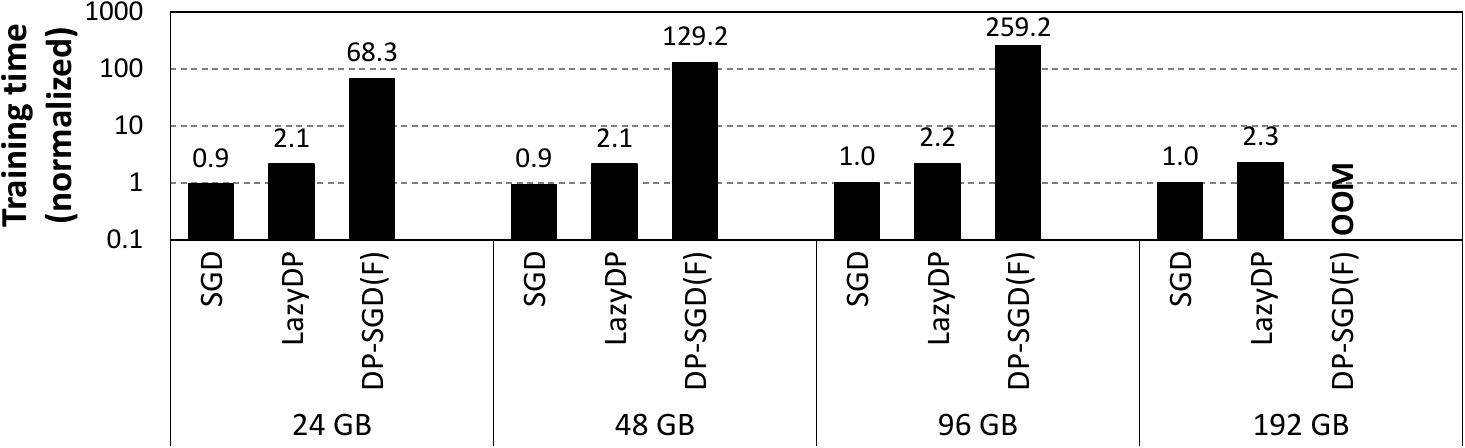}}
    \hfill
    \subfloat[]{\includegraphics[width=0.47\textwidth]{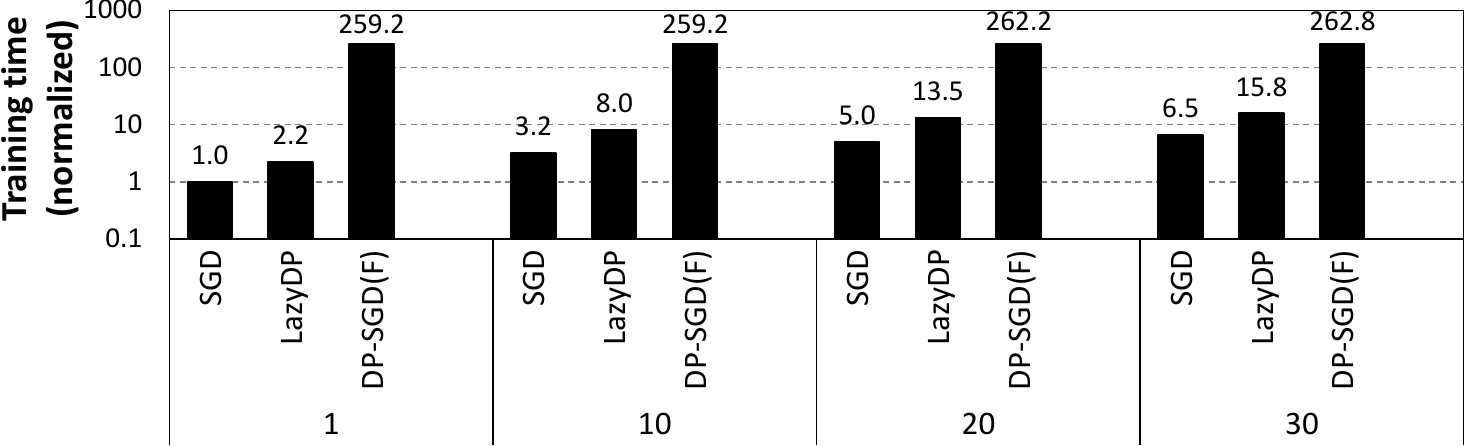}}
    \hfill
    \subfloat[]{\includegraphics[width=0.47\textwidth]{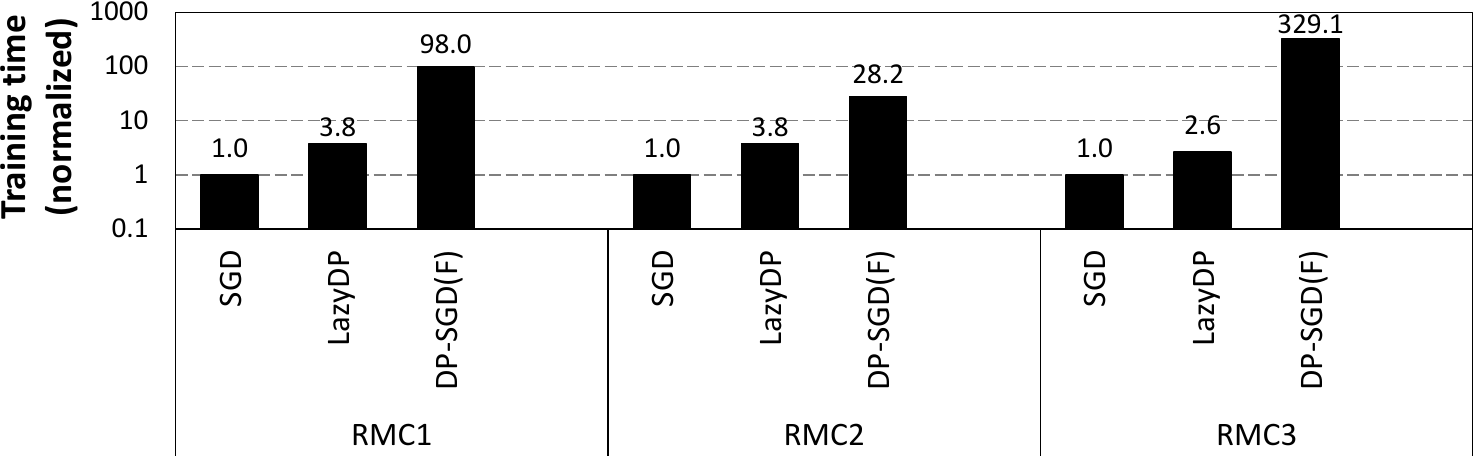}}
    
    \vspace{-0.5em}
    \hfill
    \subfloat[]{\includegraphics[width=0.47\textwidth]{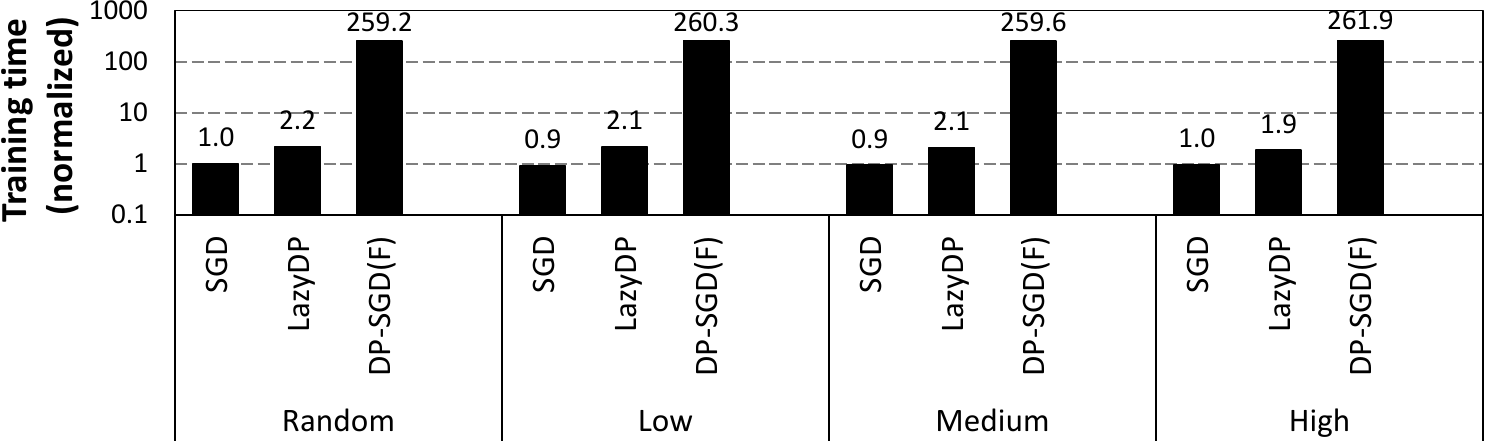}}
    \caption{\proposed sensitivity to (a) embedding table size (24/48/96/192 GB), (b) number of embedding gathers per table, i.e., the embedding pooling factor (1/10/20/30), (c) alternative DLRM model configurations, and (d) alternative training dataset.}
    
    \label{fig:sensitivity}
   \vspace{-1em}
\end{figure}

\subsection{Sensitivity}
\label{sect:eval:sensitivity}

{\bf Embedding table size.} \fig{fig:sensitivity}(a) compares the training time of \proposed against SGD and \dpsgdf when the embedding table size is changed.  Regardless of the table size, the latency of \sgd and \proposed remains almost constant. This is because the performance of \sgd and \lazydp is mostly determined as a function of the number of embedding vectors gathered from the table, irrespective of the table size. The training time of \dpsgdf, on the other hand, scales proportional to the embedding table size, causing an out-of-memory (OOM) error when the table size reaches $192$ GB. These results highlight the high scalability of \proposed, demonstrating its applicability for future large-scale RecSys models with even larger embedding table sizes.

{\bf Embedding pooling value.} 
\fig{fig:sensitivity}(b) summarizes the effect of the number of embedding vector gathers (i.e., embedding pooling value) on training time (normalized to the leftmost SGD design). Due to the large overhead incurred with its noisy model updates, the latency of \dpsgdf is already significantly high that the additional overhead caused by having a larger embedding pooling value becomes  marginal beyond pooling value of $10$. As for \sgd and \proposed, the effect of noisy model update on performance is non-existent (\sgd) or close to zero (\proposed). This means that a larger  pooling value causes a higher memory bandwidth pressure and accordingly, a higher latency overhead for \sgd and \proposed, which explains its gradual increase in training time with larger pooling values. Consequently, the performance gap between \proposed and \dpsgdf becomes narrower as the pooling value is increased. Nonetheless, \proposed still provides significant benefits and achieves $16.7\times$ speedup vs. \dpsgdf with pooling value of $30$.

{\bf Alternative DLRM model configurations.}
To further demonstrate the robustness of \proposed, we examine the effect of changing both embedding table size and embedding pooling value  by studying different DLRM model configurations discussed in \cite{deeprecsys,dlrm:arch}. As depicted in \fig{fig:sensitivity}(c), \proposed consistently provides a significant $52.7\times$ average speedup compared to \dpsgdf. With this new model architecture, \proposed incurs less than $3.1\%$ memory capacity overhead across all studied models. As for the performance overhead, \proposed introduces latency to identify which embeddings to apply noisy updates and etc, which accounts for $8.9\%$ to $11.9\%$ of \proposed's end-to-end training time.

{\bf Alternative training dataset.}
We also study the robustness of \proposed to alternative training dataset by using the Kaggle DAC dataset~\cite{criteo:dataset}. Specifically, we follow the methodology from \cite{scratchpipe} and construct three distinct datasets where each exhibits low, medium, and high embedding table access skewness, i.e., $90\%$ of the embedding table accesses are concentrated on $36\%$, $10\%$, and $0.6\%$ of table entries for our low, medium, and high skeweness dataset. As illustrated in \fig{fig:sensitivity}(d), \dpsgdf exhibits minimal variation in its training time because its end-to-end training time is completely bottlenecked by the model update stage, suffering from significant slowdown irrespective of table access locality. \proposed again shows robustness and achieves an average $129.03\times$ speedup vs. \dpsgdf. There is no additional memory capacity overhead with this new training dataset as it is primarily determined by the model configuration. In terms of performance overhead, the additional latency introduced with \proposed is shown to be always below $14\%$ of \proposed's end-to-end training time, similar to the default training configuration.

\begin{figure}[t!] \centering
\includegraphics[width=0.47\textwidth]{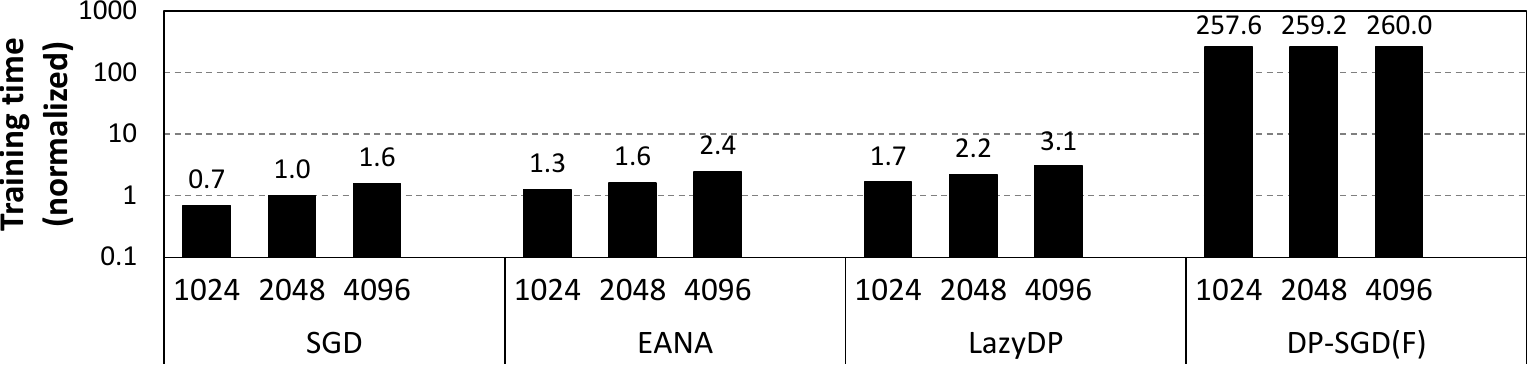}
\caption{Training time of \proposed vs. \eana over different training mini-batch sizes. All bars are normalized to SGD trained with mini-batch size 2,048.}
\label{fig:eval_eana}
\end{figure}

\subsection{LazyDP vs. EANA}
\label{sect:eval:eana}

\eana~\cite{eana} is a recently proposed private training algorithm for RecSys, presenting an alternative DP-SGD algorithm for higher performance. \eana alleviates the performance overhead of private embedding table training by adding noise \emph{only} to the accessed embedding vectors in each training iteration. While such modification helps improve training throughput, 
the privacy of \eana is highly \emph{data-dependent} and becomes 
weaker when the embedding table access pattern exhibits a skewed access distribution~\cite{eana}.
Prior work~\cite{scratchpipe,merci,recnmp,yonsei_space, ttrec} reports that the embedding table access pattern in RecSys exhibits a power-law distribution (i.e., a small number of
hot embedding vectors receive very high access frequency while the other cold vectors receive a small number of accesses), and EANA becomes less useful for such real world RecSys models. 
Overall, our \proposed is  designed to satisfy the dual requirements of high performance and privacy guarantee from the ground up. 
We implement EANA and compare against \proposed in \fig{fig:eval_eana}. \proposed only incurs $27\%$ to $37\%$ performance overhead over \eana while guaranteeing the same level of privacy provided with the vanilla DP-SGD algorithm.
\section {Conclusion}
\label{sect:conclusion}

This paper proposes \proposed,  our algorithm-software co-design that provides scalable and high-performance private RecSys training. We first characterize real world RecSys models and uncover several critical performance bottlenecks such as the compute-limited noise sampling and memory-limited noisy gradient update operations. \proposed successfully addresses these two bottlenecks via our lazy noise update and aggregated noise sampling optimizations, providing substantial performance speedup while ensuring mathematically equivalent, differentially private RecSys models to be trained.

\section*{Acknowledgment}

This research is supported by the National Research Foundation of Korea (NRF) grant funded by the Korea government (MSIT) (NRF-2021R1A2C2091753), Samsung Advanced Institute of Technology of Samsung Electronics Co., Ltd, and Samsung Electronics Co., Ltd (IO201210-07974-01). Minsoo Rhu is partially supported by the Google Research Scholar Award.

\bibliographystyle{plain}
\bibliography{refs}

\begin{thebibliography}{10}

\bibitem{abadi}
Martin Abadi, Andy Chu, Ian Goodfellow, H.~Brendan McMahan, Ilya Mironov, Kunal Talwar, and Li~Zhang.
\newblock {Deep Learning with Differential Privacy}.
\newblock In {\em Proceedings of the ACM SIGSAC Conference on Computer and Communications Security (CCS)}, 2016.

\bibitem{anil2021largescale}
Rohan Anil, Badih Ghazi, Vineet Gupta, Ravi Kumar, and Pasin Manurangsi.
\newblock {Large-Scale Differentially Private BERT}.
\newblock In {\em {arxiv.org}}, 2021.

\bibitem{apple}
Apple.
\newblock {Learning with Privacy at Scale}.
\newblock \url{https://docs-assets.developer.apple.com/ml-research/papers/learning-with-privacy-at-scale.pdf}, 2017.

\bibitem{netflix:rec}
Moumita Bhattacharya and Sudarshan Lamkhede.
\newblock {Augmenting Netflix Search with In-Session Adapted Recommendations}.
\newblock In {\em Proceedings of the ACM Conference on Recommender Systems (RECSYS)}, 2022.

\bibitem{openmp}
OpenMP Architecture~Review Board.
\newblock {OpenMP 4.5 API C/C++ Syntax Reference Guide}.
\newblock \url{https://www.openmp.org/wp-content/uploads/OpenMP-4.5-1115-CPP-web.pdf}, 2015.

\bibitem{google}
Nicholas Carlini.
\newblock {Privacy Considerations in Large Language Models}.
\newblock \url{https://ai.googleblog.com/2020/12/privacy-considerations-in-large.html}, 2020.

\bibitem{carlini_diffusion}
Nicholas Carlini, Jamie Hayes, Milad Nasr, Matthew Jagielski, Vikash Sehwag, Florian Tram{\`{e}}r, Borja Balle, Daphne Ippolito, and Eric Wallace.
\newblock {Extracting Training Data from Diffusion Models}.
\newblock In {\em {arxiv.org}}, 2023.

\bibitem{carlini_llm}
Nicholas Carlini, Florian Tram{\`{e}}r, Eric Wallace, Matthew Jagielski, Ariel Herbert{-}Voss, Katherine Lee, Adam Roberts, Tom~B. Brown, Dawn Song, {\'{U}}lfar Erlingsson, Alina Oprea, and Colin Raffel.
\newblock {Extracting Training Data from Large Language Models}.
\newblock In {\em Proceedings of the USENIX Security Symposium}, 2021.

\bibitem{wideanddeep}
Heng-Tze Cheng, Levent Koc, Jeremiah Harmsen, Tal Shaked, Tushar Chandra, Hrishi Aradhye, Glen Anderson, Greg Corrado, Wei Chai, Mustafa Ispir, Rohan Anil, Zakaria Haque, Lichan Hong, Vihan Jain, Xiaobing Liu, and Hemal Shah.
\newblock {Wide \& Deep Learning for Recommender Systems}.
\newblock In {\em Proceedings of the 1st Workshop on Deep Learning for Recommender Systems}, 2016.

\bibitem{label_only_attack}
Christopher~A. Choquette{-}Choo, Florian Tram{\`{e}}r, Nicholas Carlini, and Nicolas Papernot.
\newblock {Label-Only Membership Inference Attacks}.
\newblock In {\em {Proceedings of the International Conference on Machine Learning (ICML)}}, 2021.

\bibitem{youtube_recsys}
Paul Covington, Jay Adams, and Emre Sargin.
\newblock {Deep Neural Networks for YouTube Recommendations}.
\newblock In {\em Proceedings of the ACM Conference on Recommender Systems (RECSYS)}, 2016.

\bibitem{de2022unlocking}
Soham De, Leonard Berrada, Jamie Hayes, Samuel~L. Smith, and Borja Balle.
\newblock {Unlocking High-Accuracy Differentially Private Image Classification through Scale}.
\newblock In {\em {arxiv.org}}, 2022.

\bibitem{fast_dp_sgd}
Carson Denison, Badih Ghazi, Pritish Kamath, Ravi Kumar, Pasin Manurangsi, Krishna~Giri Narra, Amer Sinha, Avinash Varadarajan, and Chiyuan Zhang.
\newblock {Private Ad Modeling with DP-SGD}.
\newblock In {\em {arxiv.org}}, 2022.

\bibitem{amazon}
Tom Diethe, Oluwaseyi Feyisetan, Borja Balle, and Thomas Drake.
\newblock {Preserving Privacy in Analyses of Textual Data}.
\newblock In {\em Proceedings of the International Conference on Web Search and Data Mining}, 2020.

\bibitem{dwork2006dp}
Cynthia Dwork.
\newblock {Differential Privacy}.
\newblock In {\em Automata, Languages and Programming}, 2006.

\bibitem{dwork2008differential}
Cynthia Dwork.
\newblock {Differential Privacy: A Survey of Results}.
\newblock In {\em Theory and Applications of Models of Computation}, 2008.

\bibitem{dwork2014algorithmic}
Cynthia Dwork and Aaron Roth.
\newblock {The Algorithmic Foundations of Differential Privacy}.
\newblock In {\em Found. Trends Theor. Comput. Sci.}, 2014.

\bibitem{sum_of_random}
Bennett Eisenberg and Rosemary Sullivan.
\newblock {Why Is the Sum of Independent Normal Random Variables Normal?}
\newblock In {\em Mathematics Magazine}, 2008.

\bibitem{microsoft_recsys}
Ali~Mamdouh Elkahky, Yang Song, and Xiaodong He.
\newblock {A Multi-View Deep Learning Approach for Cross Domain User Modeling in Recommendation Systems}.
\newblock In {\em Proceedings of the International Conference on World Wide Web (WWW)}, 2015.

\bibitem{fb:zion}
Facebook.
\newblock {Accelerating Facebook's Infrastructure with Application-specific Hardware}.
\newblock \url{https://code.fb.com/data-center-engineering/accelerating-infrastructure/}, 2019.

\bibitem{dp_iteration}
Vitaly Feldman, Ilya Mironov, Kunal Talwar, and Abhradeep Thakurta.
\newblock {Privacy Amplification by Iteration}.
\newblock In {\em {IEEE Annual Symposium on Foundations of Computer Science (FOCS)}}, 2018.

\bibitem{fredrikson_face}
Matt Fredrikson, Somesh Jha, and Thomas Ristenpart.
\newblock {Model Inversion Attacks that Exploit Confidence Information and Basic Countermeasures}.
\newblock In {\em Proceedings of the ACM SIGSAC Conference on Computer and Communications Security (CCS)}, 2015.

\bibitem{fredrikson_drug}
Matthew Fredrikson, Eric Lantz, Somesh Jha, Simon~M. Lin, David Page, and Thomas Ristenpart.
\newblock {Privacy in Pharmacogenetics: An End-to-End Case Study of Personalized Warfarin Dosing}.
\newblock In {\em Proceedings of the USENIX Security Symposium}, 2014.

\bibitem{plug_in_inversion}
Amin Ghiasi, Hamid Kazemi, Steven Reich, Chen Zhu, Micah Goldblum, and Tom Goldstein.
\newblock {Plug-In Inversion: Model-Agnostic Inversion for Vision with Data Augmentations}.
\newblock In {\em {Proceedings of the International Conference on Machine Learning (ICML)}}, 2022.

\bibitem{amazon_mixed_dp}
Aditya Golatkar, Alessandro Achille, Yu-Xiang Wang, Aaron Roth, Michael Kearns, and Stefano Soatto.
\newblock {Mixed Differential Privacy in Computer Vision}.
\newblock In {\em Proceedings of the Conference on Computer Vision and Pattern Recognition (CVPR)}, 2022.

\bibitem{deeprecsys}
Udit Gupta, Samuel Hsia, Vikram Saraph, Xiaodong Wang, Brandon Reagen, Gu-Yeon Wei, Hsien-Hsin~S Lee, David Brooks, and Carole-Jean Wu.
\newblock {DeepRecSys: A System for Optimizing End-To-End At-Scale Neural Recommendation Inference}.
\newblock In {\em Proceedings of the International Symposium on Computer Architecture (ISCA)}, 2020.

\bibitem{dlrm:arch}
Udit Gupta, Carole-Jean Wu, Xiaodong Wang, Maxim Naumov, Brandon Reagen, David Brooks, Bradford Cottel, Kim Hazelwood, Mark Hempstead, Bill Jia, Hsien-Hsin~S. Lee, Andrey Malevich, Dheevatsa Mudigere, Mikhail Smelyanskiy, Liang Xiong, and Xuan Zhang.
\newblock {The Architectural Implications of Facebook's DNN-Based Personalized Recommendation}.
\newblock In {\em Proceedings of the International Symposium on High-Performance Computer Architecture (HPCA)}, 2020.

\bibitem{haim_reconstruction}
Niv Haim, Gal Vardi, Gilad Yehudai, Ohad Shamir, and Michal Irani.
\newblock {Reconstructing Training Data From Trained Neural Networks}.
\newblock In {\em Proceedings of the International Conference on Neural Information Processing Systems (NIPS)}, 2022.

\bibitem{hoory2021learning}
Shlomo Hoory, Amir Feder, Avichai Tendler, Sofia Erell, Alon Cohen, Itay Laish, Hootan Nakhost, Uri Stemmer, Ayelet Benjamini, Avinatan Hassidim, and Yossi Matias.
\newblock {Learning and Evaluating a Differentially Private Pre-trained Language Model}.
\newblock In {\em Findings of the Association for Computational Linguistics: EMNLP}, 2021.

\bibitem{tbb}
Intel.
\newblock {Intel oneAPI Threading Building Blocks}.
\newblock \url{https://software.intel.com/en-us/tbb}, 2023.

\bibitem{intel_pcm}
Intel.
\newblock {Intel Performance Counter Monitor (Intel PCM)}.
\newblock \url{https://github.com/intel/pcm}, 2023.

\bibitem{avx2}
Intel.
\newblock {Intrinsics for Intel Advanced Vector Extensions 2 (Intel AVX2)}.
\newblock \url{https://www.intel.com/content/www/us/en/docs/cpp-compiler/developer-guide-reference/2021-10/intrinsics-for-avx2.html}, 2023.

\bibitem{criteo:dataset}
Kaggle.
\newblock {Criteo Display Advertising Challenge}.
\newblock \url{https://www.kaggle.com/c/criteo-display-ad-challenge}, 2014.

\bibitem{yonsei_space}
Hongju Kal, Seokmin Lee, Gun Ko, and Won~Woo Ro.
\newblock {SPACE: Locality-aware Processing in Heterogeneous Memory for Personalized Recommendations}.
\newblock In {\em Proceedings of the International Symposium on Computer Architecture (ISCA)}, 2021.

\bibitem{recnmp}
Liu Ke, Udit Gupta, Carole-Jean Wu, Benjamin~Youngjae Cho, Mark Hempstead, Brandon Reagen, Xuan Zhang, David Brooks, Vikas Chandra, Utku Diril, et~al.
\newblock {RecNMP: Accelerating Personalized Recommendation with Near-Memory Processing}.
\newblock In {\em Proceedings of the International Symposium on Computer Architecture (ISCA)}, 2020.

\bibitem{kurakin2022training}
Alexey Kurakin, Shuang Song, Steve Chien, Roxana Geambasu, Andreas Terzis, and Abhradeep Thakurta.
\newblock {Toward Training at ImageNet Scale with Differential Privacy}.
\newblock In {\em {arxiv.org}}, 2022.

\bibitem{tensorcasting}
Youngeun Kwon, Yunjae Lee, and Minsoo Rhu.
\newblock {Tensor Casting: Co-Designing Algorithm-Architecture for Personalized Recommendation Training}.
\newblock In {\em Proceedings of the International Symposium on High-Performance Computer Architecture (HPCA)}, 2021.

\bibitem{scratchpipe}
Youngeun Kwon and Minsoo Rhu.
\newblock {Training Personalized Recommendation Systems from (GPU) Scratch: Look Forward Not Backwards}.
\newblock In {\em Proceedings of the International Symposium on Computer Architecture (ISCA)}, 2022.

\bibitem{sgd}
Yann LeCun, L{\'e}on Bottou, Yoshua Bengio, and Patrick Haffner.
\newblock {Gradient-Based Learning Applied to Document Recognition}.
\newblock In {\em Proceedings of the IEEE}, 1998.

\bibitem{reweighted}
Jaewoo Lee and Daniel Kifer.
\newblock {Scaling up Differentially Private Deep Learning with Fast Per-Example Gradient Clipping}.
\newblock In {\em {Proceedings on Privacy Enhancing Technologies (PoPET)}}, 2021.

\bibitem{merci}
Yejin Lee, Seong~Hoon Seo, Hyunji Choi, Hyoung~Uk Sul, Soosung Kim, Jae~W Lee, and Tae~Jun Ham.
\newblock {MERCI: Efficient Embedding Reduction on Commodity Hardware Via Sub-query Memoization}.
\newblock In {\em Proceedings of the International Conference on Architectural Support for Programming Languages and Operation Systems (ASPLOS)}, 2021.

\bibitem{li2022large}
Xuechen Li, Florian Tramer, Percy Liang, and Tatsunori Hashimoto.
\newblock {Large Language Models Can Be Strong Differentially Private Learners}.
\newblock In {\em {Proceedings of the International Conference on Learning Representations (ICLR)}}, 2022.

\bibitem{parameterhub}
Liang Luo, Jacob Nelson, Luis Ceze, Amar Phanishayee, and Arvind Krishnamurthy.
\newblock {Parameter Hub: a Rack-Scale Parameter Server for Distributed Deep Neural Network Training}.
\newblock In {\em {ACM Symposium on Cloud Computing (SoCC)}}, 2018.

\bibitem{fl}
Brendan McMahan, Eider Moore, Daniel Ramage, Seth Hampson, and Blaise~Aguera y~Arcas.
\newblock {Communication-Efficient Learning of Deep Networks from Decentralized Data}.
\newblock In {\em {Proceedings of the International Conference on Artificial Intelligence and Statistics (AISTATS)}}, 2017.

\bibitem{mlperf}
MLCommons.
\newblock {MLPerf v2.1 Training Benchmarks}.
\newblock \url{https://mlcommons.org/en/training-normal-21/}, 2022.

\bibitem{isca2022:mudigere}
Dheevatsa Mudigere, Yuchen Hao, Jianyu Huang, Zhihao Jia, Andrew Tulloch, Srinivas Sridharan, Xing Liu, Mustafa Ozdal, Jade Nie, Jongsoo Park, Liang Luo, Jie~(Amy) Yang, Leon Gao, Dmytro Ivchenko, Aarti Basant, Yuxi Hu, Jiyan Yang, Ehsan~K. Ardestani, Xiaodong Wang, Rakesh Komuravelli, Ching-Hsiang Chu, Serhat Yilmaz, Huayu Li, Jiyuan Qian, Zhuobo Feng, Yinbin Ma, Junjie Yang, Ellie Wen, Hong Li, Lin Yang, Chonglin Sun, Whitney Zhao, Dimitry Melts, Krishna Dhulipala, KR~Kishore, Tyler Graf, Assaf Eisenman, Kiran~Kumar Matam, Adi Gangidi, Guoqiang~Jerry Chen, Manoj Krishnan, Avinash Nayak, Krishnakumar Nair, Bharath Muthiah, Mahmoud khorashadi, Pallab Bhattacharya, Petr Lapukhov, Maxim Naumov, Ajit Mathews, Lin Qiao, Mikhail Smelyanskiy, Bill Jia, and Vijay Rao.
\newblock {Software-Hardware Co-Design for Fast and Scalable Training of Deep Learning Recommendation Models}.
\newblock In {\em Proceedings of the International Symposium on Computer Architecture (ISCA)}, 2022.

\bibitem{mudigere2021high}
Dheevatsa Mudigere, Yuchen Hao, Jianyu Huang, Andrew Tulloch, Srinivas Sridharan, Xing Liu, Mustafa Ozdal, Jade Nie, Jongsoo Park, Liang Luo, et~al.
\newblock {High-performance, Distributed Training of Large-scale Deep Learning Recommendation Models}.
\newblock In {\em {arxiv.org}}, 2021.

\bibitem{carlini_chatgpt}
Milad Nasr, Nicholas Carlini, Jonathan Hayase, Matthew Jagielski, A.~Feder Cooper, Daphne Ippolito, Christopher~A. Choquette-Choo, Eric Wallace, Florian Tramèr, and Katherine Lee.
\newblock {Scalable Extraction of Training Data from (Production) Language Models}.
\newblock In {\em {arxiv.org}}, 2023.

\bibitem{milad_attack}
Milad Nasr, Reza Shokri, and Amir Houmansadr.
\newblock {Comprehensive Privacy Analysis of Deep Learning: Passive and Active White-box Inference Attacks against Centralized and Federated Learning}.
\newblock In {\em {IEEE Symposium on Security and Privacy (SP)}}, 2019.

\bibitem{adversary_instantiation}
Milad Nasr, Shuang Song, Abhradeep Thakurta, Nicolas Papernot, and Nicholas Carlini.
\newblock {Adversary Instantiation: Lower Bounds for Differentially Private Machine Learning}.
\newblock In {\em {IEEE Symposium on Security and Privacy (SP)}}, 2021.

\bibitem{facebook_dlrm}
Maxim Naumov, Dheevatsa Mudigere, Hao-Jun~Michael Shi, Jianyu Huang, Narayanan Sundaraman, Jongsoo Park, Xiaodong Wang, Udit Gupta, Carole-Jean Wu, Alisson~G. Azzolini, Dmytro Dzhulgakov, Andrey Mallevich, Ilia Cherniavskii, Yinghai Lu, Raghuraman Krishnamoorthi, Ansha Yu, Volodymyr Kondratenko, Stephanie Pereira, Xianjie Chen, Wenlin Chen, Vijay Rao, Bill Jia, Liang Xiong, and Misha Smelyanskiy.
\newblock {Deep Learning Recommendation Model for Personalization and Recommendation Systems}.
\newblock In {\em {arxiv.org}}, 2019.

\bibitem{eana}
Lin Ning, Steve Chien, Shuang Song, Mei Chen, Yunqi Xue, and Devora Berlowitz.
\newblock {EANA: Reducing Privacy Risk on Large-Scale Recommendation Models}.
\newblock In {\em Proceedings of the ACM Conference on Recommender Systems (RECSYS)}, 2022.

\bibitem{nvidia_smi}
NVIDIA.
\newblock {System Management Interface SMI}.
\newblock \url{https://developer.nvidia.com/nvidia-system-management-interface}, 2016.

\bibitem{diva}
Beomsik Park, Ranggi Hwang, Dongho Yoon, Yoonhyuk Choi, and Minsoo Rhu.
\newblock {DiVa: An Accelerator for Differentially Private Machine Learning}.
\newblock In {\em Proceedings of the International Symposium on Microarchitecture (MICRO)}, 2022.

\bibitem{ponomareva-etal-2022-training-text}
Natalia Ponomareva, Jasmijn Bastings, and Sergei Vassilvitskii.
\newblock {Training Text-to-Text Transformers with Privacy Guarantees}.
\newblock In {\em Proceedings of the Fourth Workshop on Privacy in Natural Language Processing}, 2022.

\bibitem{torch}
PyTorch.
\newblock \url{http://pytorch.org}, 2022.

\bibitem{laoram}
Rachit Rajat, Yongqin Wang, and Murali Annavaram.
\newblock {LAORAM: A Look Ahead ORAM Architecture for Training Large Embedding Tables}.
\newblock In {\em Proceedings of the International Symposium on Computer Architecture (ISCA)}, 2023.

\bibitem{switchml}
Amedeo Sapio, Marco Canini, Chen-Yu Ho, Jacob Nelson, Panos Kalnis, Changhoon Kim, Arvind Krishnamurthy, Masoud Moshref, Dan Ports, and Peter Richt{\'a}rik.
\newblock {Scaling Distributed Machine Learning with In-Network Aggregation}.
\newblock In {\em {USENIX Symposium on Networked Systems Design and Implementation (NSDI)}}, 2021.

\bibitem{meta_fl_dp}
Branislav Stojkovic, Jonathan Woodbridge, Zhihan Fang, Jerry Cai, Andrey Petrov, Sathya Iyer, Daoyu Huang, Patrick Yau, Arvind Kumar, Hitesh Jawa, and Anamita Guha.
\newblock {Applied Federated Learning: Architectural Design for Robust and Efficient Learning in Privacy Aware Settings}.
\newblock In {\em {arxiv.org}}, 2022.

\bibitem{plug_n_play}
Lukas Struppek, Dominik Hintersdorf, Antonio De~Almeida Correia, Antonia Adler, and Kristian Kersting.
\newblock {Plug {\&} Play Attacks: Towards Robust and Flexible Model Inversion Attacks}.
\newblock In {\em {Proceedings of the International Conference on Machine Learning (ICML)}}, 2022.

\bibitem{chuan_attack}
Lauren Watson, Chuan Guo, Graham Cormode, and Alexandre Sablayrolles.
\newblock {On the Importance of Difficulty Calibration in Membership Inference Attacks}.
\newblock In {\em {Proceedings of the International Conference on Learning Representations (ICLR)}}, 2022.

\bibitem{box_muller_wiki}
Wikipedia.
\newblock {Box–Muller Transform}.
\newblock \url{https://en.wikipedia.org/wiki/Box-Muller_transform}, 2023.

\bibitem{yi2018factorized}
Xinyang Yi, Yi-Fan Chen, Sukriti Ramesh, Vinu Rajashekhar, Lichan Hong, Noah Fiedel, Nandini Seshadri, Lukasz Heldt, Xiang Wu, and EH~Chi.
\newblock {Factorized Deep Retrieval and Distributed TensorFlow Serving}.
\newblock In {\em {Proceedings of Conference on Machine Learning and Systems (MLSys)}}, 2018.

\bibitem{ttrec}
Chunxing Yin, Bilge Acun, Carole-Jean Wu, and Xing Liu.
\newblock {TT-Rec: Tensor Train Compression for Deep Learning Recommendation Models}.
\newblock In {\em {Proceedings of Conference on Machine Learning and Systems (MLSys)}}, 2021.

\bibitem{opacus}
Ashkan Yousefpour, Igor Shilov, Alexandre Sablayrolles, Davide Testuggine, Karthik Prasad, Mani Malek, John Nguyen, Sayan Ghosh, Akash Bharadwaj, Jessica Zhao, Graham Cormode, and Ilya Mironov.
\newblock {Opacus: {U}ser-Friendly Differential Privacy Library in {PyTorch}}.
\newblock In {\em {arxiv.org}}, 2021.

\bibitem{yu2022differentially}
Da~Yu, Saurabh Naik, Arturs Backurs, Sivakanth Gopi, Huseyin~A Inan, Gautam Kamath, Janardhan Kulkarni, Yin~Tat Lee, Andre Manoel, Lukas Wutschitz, Sergey Yekhanin, and Huishuai Zhang.
\newblock {Differentially Private Fine-tuning of Language Models}.
\newblock In {\em {Proceedings of the International Conference on Learning Representations (ICLR)}}, 2022.

\bibitem{aibox}
Weijie Zhao, Jingyuan Zhang, Deping Xie, Yulei Qian, Ronglai Jia, and Ping Li.
\newblock {AIBox: CTR Prediction Model Training on a Single Node}.
\newblock In {\em Proceedings of the ACM International Conference on Information and Knowledge Management}, 2019.

\end{thebibliography}

\end{document}